\documentclass[12pt,a4paper]{article}
\pdfoutput=1
\usepackage[format=hang]{caption}
\usepackage{epsfig,amssymb,amsmath,graphicx,subcaption,verbatim,hyperref,xcolor,ulem,sidecap,bm,float,tensor,comment}
\newcommand{\be}{\begin{equation}}
\newcommand{\ee}{\end{equation}}

\newcommand{\news}{\setcounter{equation}{0}}
\def\bea{\begin{eqnarray}}
\def\eea{\end{eqnarray}}
\numberwithin{equation}{section}
\makeatletter
\renewcommand*\env@matrix[1][\arraystretch]{
  \edef\arraystretch{#1}
  \hskip -\arraycolsep
  \let\@ifnextchar\new@ifnextchar
  \array{*\c@MaxMatrixCols c}}
\makeatother

\begin{document}
\title{\vskip -65pt
\begin{flushright}
\end{flushright}
\vskip 60pt
{\bf {\LARGE Gauge choices and Entanglement Entropy of two dimensional lattice gauge fields}}\\[20pt]}

\author{\bf {\Large  Zhi Yang$^{a}$\footnote{zyang14@fudan.edu.cn}  \ and Ling-Yan Hung$^{a,b}$\footnote{elektron.janethung@gmail.com}} \\[25pt]
\it ${}^{a}$ Department of Physics and Center for Field Theory and Particle Physics,\\ Fudan University,
220 Handan Road, 200433 Shanghai, China\\
\it ${}^{b}$State Key Laboratory of Surface Physics and Department of Physics,\\ Fudan University,
220 Handan Road, 200433 Shanghai, China\\
}

\date{\today}
\maketitle
\vskip 20pt

\begin{abstract}

In this paper, we explore the question of how different gauge choices in a gauge theory affect the tensor product structure of the Hilbert space in configuration space. In particular, we study the Coulomb gauge and observe that the naive gauge potential degrees of freedom cease to be local operators as soon as we impose the Dirac brackets. We construct new local set of operators and compute the entanglement entropy according to this algebra in $2+1$ dimensions. We find that our proposal would lead to an entanglement entropy that behave very similar to a single scalar degree of freedom if we do not include further centers, but approaches that of a gauge field if we include non-trivial centers. We explore also the situation where the gauge field is Higgsed, and construct a local operator algebra that again requires some deformation. This should give us some insight into interpreting the entanglement entropy in generic gauge theories and perhaps also in gravitational theories.

\end{abstract}

\vskip 20pt

PACS:
\newpage

\section{Introduction}\news

The seminal papers \cite{Buividovich:2008gq,Casini:2013rba} initiated the important question of defining the notion of entanglement entropy of a gauge theory.  The definition of the entanglement entropy has been heavily based on a tensor product structure in the Hilbert space. To study entanglement in configuration space, it requires a tensor product structure in configuration space. Naively, this is a natural feature in the Hilbert space, since the world we experienced around us is local, and that it gives the impression that the Hilbert space naturally factorizes as a tensor product of spaces defined locally at each point in space. There are clear subtlties when we work with gauge theories, where it is well known that gauge theories are by construction made local by including gauge degrees of freedom. The gauge constraints such as the Gauss law implies that degrees of freedom at different locations are not entirely independent, and thus a naive factorization of the Hilbert space is not possible. This is clearly a significant issue both for gauge theories and gravitational theories \cite{Donnelly:2016auv, Donnelly:2016rvo}. A clear understanding is thus crucial, also in ultimately formulating a theory of quantum gravity.

Until very recently, the replica trick has been the main tool employed to computing the entanglement enropy or the Renyi entropy in field theories, which in turn can be formulated as a path-integral in a conical space. This allows one to momentarily brush off issues of the Hilbert space and obtain some results -- until it is realized that the issue in fact re-emerge as some ambiguities with edge modes that are localized at the entangling surface that is only recently understood \cite{Donnelly:2014fua,Donnelly:2015hxa}. See also \cite{ Huang:2016bkp, Ma:2015xes}.

Since the introduction of the notion of center in  \cite{Casini:2013rba} into the discussion of entanglement entropy, it has made sense of entanglement even as the Hilbert space does not admit a factorization. Instead of directly considering the Hilbert space, one formulates the question in terms of the choice of an algebra attached to some region. There is some ambiguity in the selection of the algebra, and they can be characterized by different centers.  A large amount of work is inspired to understand the physical significance of these difference choices \cite{Soni:2015yga,Ghosh:2015iwa,Soni:2016ogt,Donnelly:2014gva,Aoki:2015bsa,Radicevic:2014kqa,Gromov:2014kia,
Hung:2015fla,Radicevic:2015sza,Mathur:2015wba,VanAcoleyen:2015ccp}, and to demonstrate that these arbitrary choices can approach the same value as the continuous limit is taken in the mutual information for example \cite{Casini:2013rba}.

The discussion in \cite{Casini:2014aia} is mainly phrased directly in terms of gauge invariant degrees of freedom. However, most lattice gauge theories or the field theoretic studies of gauge theories are formulated in terms of the gauge potential, and a Fock space is constructed for the gauge potential. Most of the discussions of entanglement in lattice gauge theories proceed by picking the temporal gauge $A_0 = 0$, and that the gauge potential lives on the links of the lattice, with gauge invariance imposed at the vertex. For example in the original paper \cite{Casini:2013rba} and most other references, this is essentially the choice. This suggests a natural question: have we exhausted all the subtleties posed by the non-locality of gauge theories? Do we understand the operator algebra and how they are attached to local regions in an arbitrary gauge? Notwithstanding the introduction of centers, it is still necessary that there is an approximate choice of local operators that can be associated to some region for entanglement in configuration space to be meaningful.

We therefore take the first step in this direction, and explore the quantization of a U(1) gauge theory in the Coulomb gauge. It is well known that by imposing the Coulomb gauge, standard Poisson brackets have to be modified by Dirac brackets to take into account the gauge condition which in this case are second class constraints. Perhaps unsurprisingly, the gauge potential and its conjugate momenta cease to be local operators. Their Dirac commutators are non-vanishing even as the operators are separated by very large distances.

To make sense of the entanglement, it becomes necessary to construct local operators. We found a suitable construction making use of the duality relation between a vector and a scalar in 2d.  We proposed a construction of local operators in terms of the now non-local gauge potentials.
As soon as a choice of a complete set of local operators have been chosen, then the computation of entanglement entropy would resemble the usual prescription. In particular, we would have a freedom to discard within this set of local operators various finite set close to the boundary and generically generate a center in the operator algebra.
The entanglement entropy can be obtained for each such choice of algebra in a straightforward manner. We note however that our choice of local operators are very different from the choice in the literature -- the natural choice of local operators in the current gauge choice is based on the gauge potentials and their time derivatives rather than the electric or magnetic fields. Particularly, taking the gauge potential $A$ as the fundamental degree of freedom, we study the case corresponding to a trivial center,  a ``field center'' where some $A$ operators are taken into the center, and the ``momentum center'', where various time derivatives of $A$ at the boundary is taken into the center. These are the natural analogous to the trivial center, the electric center/magnetic center respectively.  However, since the fundamental degree of freedom is different, they generically mean something different from the discussion of trivial/electric/magnetic centers that appear in the literature.

In any event, It appears that the proposal is robust - it is insensitive to various local prescriptions at the boundary.  However the log term associated to corners seem to approach that of a local scalar field in the case of a trivial center, but approaches that of a gauge field when we pick a center mimicking the electric/magnetic center.

Then we consider a U(1) theory in the Higgsed phase, and apply our prescription still picking the Coulomb gauge. We recover a mutual information that decays exponentially according to the ratio of the mass and the lattice scale.  A gauge theory coupled to matter in 1+1 d has been considered in \cite{Aoki:2017ntc}, although the gauge theory has no dynamical degrees of freedom and the only remnant is Gauss's law. Here we have an example in which matter interacts with dynamical gauge fields.

Our paper is organized as follows.  In section 2 we review \cite{ Casini:2014aia} briefly the computation of entanglement entropy of a quadratic theory making use of  the values of the correlators and commutators. In section 3, we discuss the quantization of the U(1) gauge field in Coulomb gauge and the associated Dirac brackets. We discretize the theory in section 4, and introduce dual scalar variables, which allowed to construct truely local operators.  In section 5, we computed the entanglement entropy based on our prescription.   Particularly we will make a detailed comparison of our choices of centers with the traditional literature. In section 6-8, we generalize our method to the Higgsed gauge theory and computed the entanglement entropy accordingly.
We look into mutual information in section 9 and finally conclude in section 10.

\section{Entropies of Gaussian States in terms of correlation functions}
In this section, we briefly review the methods described in \cite{Casini:2014aia}. It is demonstrated that  expressions for entanglement entropy can be readily expressed in terms of commutators and correlations of some (canonical) variables in a quadratic theory.\\
We consider the general commutation relations
\be\label{Cij}
[q_i,p_j]=i C_{ij},
\ee
and correlation functions
\be
\langle p_i,p_j\rangle=P_{ij},
\ee
\be
\langle q_i,q_j\rangle=X_{ij},
\ee
\be
\langle q_i,p_j\rangle=\frac{i}{2}C_{ij},
\ee
with $i,j\in V$. In the trivial center case, we have the entanglement entropy of region $V$ \cite{Casini:2014aia}
\be
S(V)=tr((\Theta +1/2)\log(\Theta +1/2)-(\Theta -1/2)\log(\Theta -1/2)),
\ee
where $\Theta=\sqrt{C^{-1}X(C^{-1})^TP}$. $X$ and $P$ are matrices of correlation functions and $C$ are matrix of commutators.\\
The method can be generalized to the case with center. We consider the algebra generated by $q_i,p_j$, with $i\in V={1,\ldots n}$ and $j\in B={k+1,\ldots n}$ with $B\subset V$. We assume $[q_i,p_j]=0$ for $i\in A={1,\ldots k}$,$j\in B$, such that $q_i,i\in A$ span the center of the algebra. In the case with center, the entanglement entropy is defined as \cite{EE}
\be
S(V)=S_Q(B)+H(A),
\ee
where $S_Q(B)$ is an average of quantum contributions and $H(A)$ is the classical Shannon entropy. With the expressions of correlation functions and commutators, we have\cite{Casini:2014aia}
\be
S_Q(B)=tr((\Theta +1/2)\log(\Theta +1/2)-(\Theta -1/2)\log(\Theta -1/2)),
\ee
\be
\Theta=\sqrt{\tilde{X}\tilde{P}},
\ee
\be
\tilde{X}=(C^T_{VB}X_V^{-1}C_{VB})^{-1}, \qquad \tilde{P}=P_B.
\ee
Here $C_{VB}$ is the commutation matrix (\ref{Cij}) between $q_i$ with $i\in V$ and $p_j$ with $j\in B$. The classical part has the form
\be
H(A)=\frac{1}{2}tr(1+\log(2\pi X_A)).
\ee
The case of center formed by $p_i$ with $i\in A={1,\ldots k}$ can be analyzed in the same way, interchanging $X\leftrightarrow P$.
\section{Local Operators of $U(1)$ Gauge Fields with Coulomb Gauge}\news
We consider the $U(1)$ gauge fields with Coulomb gauge in $2+1$ dimensions. The Lagrangian is
\be
\mathcal{L}=-\frac{1}{4}F_{\mu\nu}F^{\mu\nu}
\ee
and the Coulomb gauge fixing is
\be\label{coulomb gauge}
\nabla \cdot \vec{A}=0.
\ee
The temporal component is non dynamical and it needs only to satisfy a constraint following from the Gauss law and the gauge constraint, relating it to the total charge.  Because we consider free gauge fields with no charged matter, we can set the temporal component $A_0$ to be 0. We have the canonical momentum
\be
\pi^i=\frac{\partial\mathcal{L}}{\partial \dot{A}_i}=\dot{A}_i.
\ee
The Gauss law and the gauge constraint also imply
\be\label{gaussian law}
\partial_i \pi^i=0.
\ee
To impose the two constraints (\ref{coulomb gauge}) and (\ref{gaussian law}), we have to consider the Dirac bracket. A detailed discussion can be found in  \cite{Weinberg}.  The commutators
\be\label{commutator}
\begin{aligned}
&[A_i(\vec{x}),\pi^j(\vec{y})]=i \delta_i^j \delta (\vec{x}-\vec{y}) + i \frac{\partial^2}{\partial x^j \partial x^i} \frac{1}{4\pi |\vec{x}-\vec{y}|}\\
=&i \delta_i^j \frac{1}{(2\pi)^2}\int d^2\vec{k} e^{i \vec{k}\cdot (\vec{x}-\vec{y})} +i  \frac{\partial^2}{\partial x^j \partial x^i}\frac{1}{(2\pi)^2}\int d^2\vec{k} \frac{e^{i \vec{k}\cdot (\vec{x}-\vec{y})}}{|\vec{k}|^2}\\
=&\frac{i}{(2\pi)^2}\int d^2\vec{k} (\delta_i^j-\frac{k_i k_j}{|\vec{k}|^2})e^{i \vec{k}\cdot (\vec{x}-\vec{y})}
\end{aligned}
\ee
and
\be\label{commutatorAA}
[A_i(\vec{x}),A_j(\vec{y})]=[\pi^i(\vec{x}),\pi^j(\vec{y})]=0.
\ee
Since we consider a 2+1 dimensional theory, there is only one degree of physical freedom in Maxwell fields, also only one polarization. The mode expansions for gauge fields and their canonical momenta are given by
\be\label{Ai}
A_i (\vec{x},t)=\frac{1}{2\pi}\int \frac{d^2k}{\sqrt{2\omega}} \left[e^{-i \omega t+i \vec{k}\cdot \vec{x}}e_i(\vec{k})\hat{a}(\vec{k},\sigma)+e^{i \omega t-i \vec{k}\cdot \vec{x}}e_i(\vec{k})^*\hat{a}^*(\vec{k},\sigma)\right]
\ee
and
\be
\pi^j(\vec{x},t)=\frac{1}{2\pi}\int \frac{d^2k}{\sqrt{2\omega}} \left[-i \omega e^{-i \omega t+i \vec{k}\cdot \vec{x}}e_i(\vec{k})\hat{a}(\vec{k},\sigma)+i \omega e^{i \omega t-i \vec{k}\cdot \vec{x}}e_i(\vec{k})^*\hat{a}^*(\vec{k},\sigma)\right]
\ee
To satisfy the commutator (\ref{commutator}), the polarization $e_i(\vec{k},\sigma)$ have to satisfy
\be\label{eiejcoulomb}
e_i(\vec{k}) e_j(\vec{k})^*=\delta_{ij}-\frac{k_i k_j}{|\vec{k}|^2}.
\ee
Interestingly, by modifying the brackets by the Dirac method, we have made some rather drastic change to the tensor product structure of the Hilbert space. The gauge potentials and their conjugate momenta, even before applying the Gauss's constraint, can no longer be considered as a local degree of freedom, i.e., the operators $A_i$ and $\pi^j$ are not local. From (\ref{commutator}), one can see that their commutators are not local. They remain non-vanishing even though the fields are separated by large distances. To discuss entanglement entropy, we need to recover a basis of local operators.  To our knowledge, we are not aware of a standard method of defining a suitable set of basis in such a situation. We therefore propose the following.  Consider the operators
\be
\hat{A}_i\equiv-\nabla^2 A_i, \quad \pi^j.
\ee
The two constraints of the new variables $\hat{A}_i$ and $\pi^j$ are
\be\label{Ahatfixing}
\nabla \cdot \vec{\hat{A}}=0
\ee
and
\be\label{gaussianlaw}
\partial_i \pi^i=0.
\ee
We find that the commutators of new operators $\hat{A}_i$ and $\pi^j$ are
\be\label{Ahatpi}
[\hat{A}_i,\pi^j]=\frac{i}{(2\pi)^2}\int d^2\vec{k} (\delta_i^j|\vec{k}|^2-k_i k_j)e^{i \vec{k}\cdot (\vec{x}-\vec{y})}
\ee
and
\be\label{commutatorAhatAhat}
[\hat{A}_i(\vec{x}),\hat{A}_j(\vec{y})]=[\pi^i(\vec{x}),\pi^j(\vec{y})]=0.
\ee
We can see that the operators $\hat{A}_i$ and $\pi^j$ are local. We will consider the duality of them in the lattice and calculate the entanglement entropy.
\section{$U(1)$ Gauge Fields with Coulomb Gauge duality in the lattice}
\begin{figure}[!h]
		\centering
		\includegraphics[width=8cm]{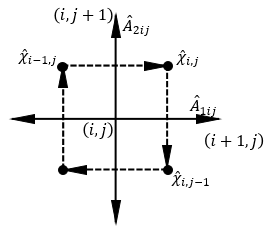}
		\caption{Dual lattice: The scalar field operator $\hat{\chi}$ is in the center of the plaquette, and the gauge field operator $\hat{A}$ in some link is equal to a difference of scalar fields across the link in the dual lattice which is perpendicular to the one corresponding to $\hat{A}$. The duality of gauge field momentum operator $\pi$ is in the same way.}
		\label{coulombdual}
\end{figure}

In (2+1) dimensional $U(1)$ gauge fields, the polarization constraint (\ref{eiejcoulomb}) implies a solution
\be
e_i(k)=i\epsilon_{ij}\frac{k_j}{\omega},
\ee
where $\omega=|\vec{k}|$ in this section. From this solution, we can see that the Maxwell fields are dual to a scalar field $\chi$ in a fixed time slice in the two dimensional spatial slice. The duality is written as
\be \label{eq:dual1}
A_i=\epsilon_{ij}\partial_j\chi,
\ee
giving the following identifications
\be
A_1=\partial_2\chi,
\ee
\be
A_2=-\partial_1\chi.
\ee
From (\ref{Ai}) and (\ref{eiejcoulomb}), we have the mode expansion of $\chi$
\be
\chi(\vec{x},t)=\frac{1}{2\pi}\int\frac{d^2\vec{k}}{\sqrt{2\omega}}\left(\frac{1}{\omega}e^{-i\omega t+i\vec{k}\cdot\vec{x}}a(\vec{k})+\frac{1}{\omega}e^{i\omega t-i\vec{k}\cdot\vec{x}}a^\dag(\vec{k})\right).
\ee
Therefore, it follows that
\be
\pi^i=\epsilon_{ij}\partial_j\dot{\chi}
\ee
with
\be\label{chidot}
\dot{\chi}(\vec{x},t)=\frac{1}{2\pi}\int\frac{d^2\vec{k}}{\sqrt{2\omega}}\left(-ie^{-i\omega t+i\vec{k}\cdot\vec{x}}a(\vec{k})+ie^{i\omega t-i\vec{k}\cdot\vec{x}}a^\dag(\vec{k})\right).
\ee
Let us define
\be\label{chihat}
\hat{\chi}(\vec{x},t)\equiv-\nabla^2\chi(\vec{x},t)=\frac{1}{2\pi}\int\frac{d^2\vec{k}}{\sqrt{2\omega}}\left(\omega e^{-i\omega t+i\vec{k}\cdot\vec{x}}a(\vec{k})+\omega e^{i\omega t-i\vec{k}\cdot\vec{x}}a^\dag(\vec{k})\right).
\ee
For the operators $\hat{\chi}$ and $\dot{\chi}$, we have the commutators
\be
[\hat{\chi}(\vec{x},t),\dot{\chi}(\vec{y},t)]=i\delta^2(\vec{x}-\vec{y})
\ee
and
\be
[\hat{\chi}(\vec{x},t),\hat{\chi}(\vec{y},t)]=[\dot{\chi}(\vec{x},t),\dot{\chi}(\vec{y},t)]=0.
\ee
For the local operators $\hat{A}_i$ and $\pi^j$, we have  similar relations
\be\label{Ahatdual}
\hat{A}_i=\epsilon_{ij}\partial_j\hat{\chi},
\ee
and
\be\label{pidual}
\pi^j=\epsilon_{jk}\partial_k\dot{\chi}.
\ee
Now we discretize the model in a square lattice. We define the operators $\hat{A}_1$ and $\pi^1$ associated to horizontal links, $\hat{A}_2$ and $\pi^2$ to vertical links, as shown in figure (\ref{coulombdual}). For example, we have $\hat{A}_{1(ij,i+1j)}$ associated to a horizontal link and $\hat{A}_{2(ij,ij+1)}$ to vertical link, where $(ij,i'j')$ are coordinates of the initial and final points of the links. For simplicity, we label them with the initial vertex of the vector,
\be
\hat{A}_{1ij}\equiv\hat{A}_{1(ij,i+1j)},
\ee
\be
\hat{A}_{2ij}\equiv\hat{A}_{2(ij,ij+1)}.
\ee
The discrete version of (\ref{Ahatdual}) and (\ref{pidual}) is also shown in figure (\ref{coulombdual}). The operators $\hat{A}$ and $\pi$ are related to the differences of the scalar field operators $\hat{\chi}$ and $\dot{\chi}$ in the orthogonal direction in the dual lattice respectively, such as
\be\label{Ahat1}
\hat{A}_{1ij}=\hat{\chi}_{i,j}-\hat{\chi}_{i,j-1},
\ee
\be\label{Ahat2}
\hat{A}_{2ij}=\hat{\chi}_{i-1,j}-\hat{\chi}_{i,j},
\ee
and
\be\label{pi1}
\pi^1_{ij}=\dot{\chi}_{i,j}-\dot{\chi}_{i,j-1},
\ee
\be\label{pi2}
\pi^2_{ij}=\dot{\chi}_{i-1,j}-\dot{\chi}_{i,j}.
\ee
Because there are redundant degrees of freedoms in gauge fields, we have two constraints (\ref{Ahatfixing}) and (\ref{gaussianlaw}) with operators $\hat{A}_i$ and $\pi^j$. In the discrete lattice, the two constraints become
\be\label{Aabfixing}
\sum_{b}\hat{A}_{ab}=0
\ee
and
\be\label{piabfixing}
\sum_{b}\pi^{ab}=0,
\ee
where the sum is over all the links $(ab)$ with the common vertex $a$. In the above equations, it is assumed that the field component is the corresponding one to the link direction. The links have orientations, which changes the field attached to it when changing the orientation, such as $\hat{A}_{ab}=-\hat{A}_{ba}$.\\
With the above dualities, the non-zero commutators of the discrete version of operators $\hat{A}$ and $\pi$ are
\be
[\hat{A}_{1ij},\pi^1_{kl}]=i(2\delta_{ik}\delta_{jl}-\delta_{ik}\delta_{j,l-1}-\delta_{ik}\delta_{j-1,l}),
\ee
\be
[\hat{A}_{1ij},\pi^2_{kl}]=i(\delta_{i,k-1}\delta_{jl}+\delta_{ik}\delta_{j-1,l}-\delta_{ik}\delta_{jl}-\delta_{i,k-1}\delta_{j-1,l}),
\ee
\be
[\hat{A}_{2ij},\pi^1_{kl}]=i(\delta_{i-1,k}\delta_{jl}+\delta_{ik}\delta_{j,l-1}-\delta_{ik}\delta_{jl}-\delta_{i-1,k}\delta_{j,l-1}),
\ee
and
\be
[\hat{A}_{2ij},\pi^2_{kl}]=i(2\delta_{ik}\delta_{jl}-\delta_{i-1,k}\delta_{j,l}-\delta_{i,k-1}\delta_{jl}).
\ee
We can see that the discrete version of operators $\hat{A}$ and $\pi$ are almost local.  We use these operators to calculate the entanglement entropy in section 5.
\\
From (\ref{chihat}) and (\ref{chidot}), the vacuum correlation functions of operators $\hat{\chi}$ and $\dot{\chi}$ are found to be
\be
\langle \hat{\chi}(\vec{x},t)\hat{\chi}(\vec{y},t)\rangle=\frac{1}{(2\pi)^2}\int d^2k \frac{\omega}{2}e^{i \vec{k}\cdot (\vec{x}-\vec{y})},
\ee
\be
\langle \dot{\chi}(\vec{x},t)\dot{\chi}(\vec{y},t)\rangle=\frac{1}{(2\pi)^2}\int d^2k \frac{1}{2\omega}e^{i \vec{k}\cdot (\vec{x}-\vec{y})}.
\ee
The vacuum correlation functions of the discrete version are
\be
\langle \hat{\chi}_{i,j}\hat{\chi}_{k,l}\rangle=\frac{1}{(2\pi)^2}\int_{-\pi}^\pi d k_x \int_{-\pi}^\pi d k_y\sqrt{\sin^2\frac{k_x}{2}+\sin^2\frac{k_y}{2}}\cos(k_x(i-k))\cos(k_y(j-l)),
\ee
\be
\langle \dot{\chi}_{i,j}\dot{\chi}_{k,l}\rangle=\frac{1}{(2\pi)^2}\int_{-\pi}^\pi d k_x \int_{-\pi}^\pi d k_y\frac{\cos(k_x(i-k))\cos(k_y(j-l))}{4\sqrt{\sin^2\frac{k_x}{2}+\sin^2\frac{k_y}{2}}}.
\ee
The vacuum correlation functions of discrete variables $\hat{A}_{ij}$ and $\pi_{ij}$ can be expressed with the above correlation functions, such as
\be
\langle \hat{A}_{1ij}\hat{A}_{1kl}\rangle=2\langle \hat{\chi}_{i,j}\hat{\chi}_{k,l}\rangle-\langle \hat{\chi}_{i,j}\hat{\chi}_{k,l-1}\rangle-\langle \hat{\chi}_{i,j-1}\hat{\chi}_{k,l}\rangle.
\ee

\section{Entanglement Entropy of two dimensional lattice gauge fields with Coulomb gauge}
\begin{figure}[!h]
		\centering
		\includegraphics[width=12cm]{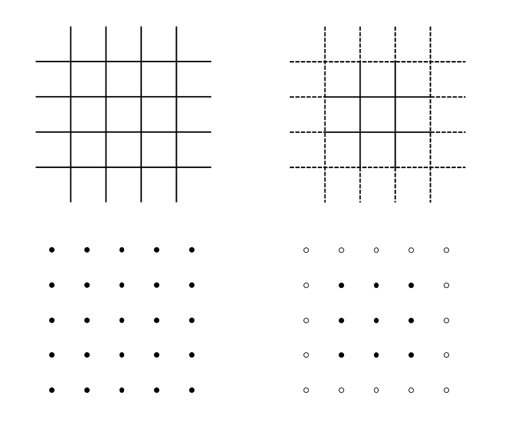}
		\caption{Gauge fields in Coulomb gauge on the lattice. The figures correspond to square regions of size $n=3$. The top two figures correspond to the gauge field and the bottom ones to the dual scalar field representation of the same algebras. Links with solid lines mean both the corresponding operators $\hat{A}$ and $\pi$ belong to the algebra. Links with dashed lines mean the corresponding operator $\hat{A}$ or $\pi$ does not belong to the algebra. Marked dots correspond to both the scalar field operators $\hat{\chi}$ and $\dot{\chi}$. Circle dots mean the scalar field operator $\hat{\chi}$ or $\dot{\chi}$ does not belong to the algebra. The left panel shows the trivial center choice, while the right panel is related to the $\hat{A}$ center or $\pi$ center choice (but not exactly), according to the meaning of dashed lines.}
		\label{latticecoulomb}
\end{figure}

\begin{figure}[!h]
		\centering
		\includegraphics[width=12cm]{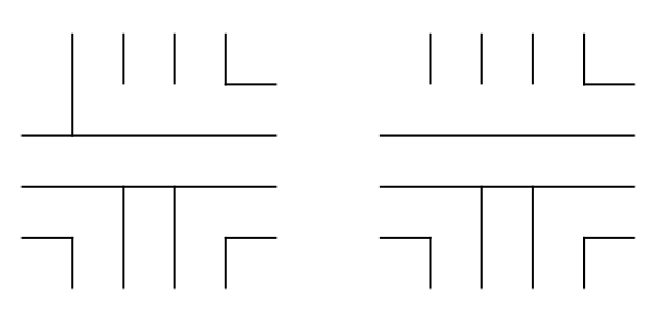}
		\caption{Two trivial centers of gauge fields with Coulomb gauge. The figures are corresponding to square regions of size $n=3$. Because of the redundancy of gauge fields, we have to fix some operators on the links to get the physical degrees of freedom. For every vertex, we can fix a link connected to it, but we can not fix a loop. We remove the fixed link in the above two figures. The left figure corresponds to the algebra of full trivial center, where we keep all the physical degree of freedom. We denote it as $\mathcal{A}$. The right figure corresponds to the algebra of trivial center with one physical degree of freedom removed. We denote it as $\hat{\mathcal{A}}$. }
		\label{coulombtrivial}
\end{figure}

\begin{figure}[!h]
		\centering
		\includegraphics[width=12cm]{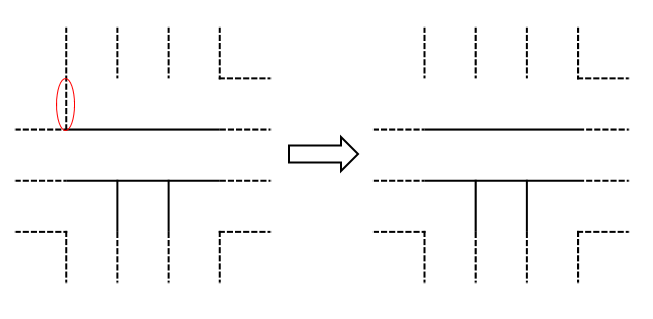}
		\caption{The non-trivial centers of gauge fields with Coulomb gauge. The figures correspond to square regions of size $n=3$. In the above two figures, we have done the gauge fixing and all the operators on the links are physical. Links with solid lines mean both the corresponding operators $\hat{A}$ and $\pi$ belong to the algebra and links with dashed lines mean the corresponding operator $\hat{A}$ or$\pi$ does not belong to the algebra. Without loss of generality, let us assume operators $\hat{A}$ to be on the dashed links, not $\pi$. In the left figure, the operator $\hat{A}$ on the dashed link in the red ellipse does not commute with $\pi$ on some dashed links. In the right figure, we remove that dashed link and all operators on dashed links form an $\hat{A}$ center. We denote such an algebra as $\mathcal{A}^{\hat{A}}$. When the dashed links correspond to operators $\pi$, not $\hat{A}$, we get the algebra of $\pi$ center. we denote it as $\mathcal{A}^{\pi}$.}
		\label{coulombcenter}
\end{figure}

To calculate the entanglement entropy of some ``region" in the lattice, we have to choose an algebra of local operators to define the ``region". In the case of gauge fields, the gauge fields operators are associated to the links. Once we have chosen the fundamental sets of local operators, we can study the entanglement entropy corresponding to different choices of algebras -- we can choose to discard from our set of local operators a subset so that there is a resultant center. We study four possible choices of algebras, which are shown in figure \ref{latticecoulomb}, figure \ref{coulombtrivial} and figure \ref{coulombcenter}. The four choices of algebras are, respectively, 1) trivial center $\mathcal{A}$, 2) trivial center with one physical degree of freedom removed $\hat{\mathcal{A}}$, 3) $\hat{A}$ center $\mathcal{A}^{\hat{A}}$ and 4)  $\pi$ center $\mathcal{A}^\pi$. In figure \ref{latticecoulomb}, we illustrate graphically the  different algebras also in terms of the gauge potential and also their dual scalar variables.
We note that these are natural analogues of the trivial center choice, ``field center'' and `` momentum center''  considered in other theories such as simple scalar field theories \cite{Casini:2013rba}, as soon as a natural choice of local degrees of freedom has been selected.
\\

Due to the redundancy of degrees of freedom in the gauge potentials, we have to remove any remaining unphysical degrees of freedom. From the two constraints (\ref{Aabfixing}) and (\ref{piabfixing}), we have to remove one degree of freedom with a vertex. That is, for every vertex, we have to remove one link connected to it. Note that it is not possible to keep all the external links of any regions, since there is an overall constraint. They are not independent.\\

In the trivial center choices, as shown in figure \ref{coulombtrivial}, both operators $\hat{A}$ and $\pi$ are associated to every link in the figure. We keep the same number  of $\hat{A}$ and $\pi$. In the left figure, we keep all the physical degrees of freedom, which is obtained by removing the unphysical degrees of freedom of gauge potential in the left panel of figure \ref{latticecoulomb}, while in the right figure, we remove a physical degree of freedom.\\

To get the $\hat{A}$ center choice, we remove the unphysical degrees of freedom of the gauge field in the right panel of figure \ref{latticecoulomb}, in which we remove all the operators $\pi$ associated to the boundary links, and then remove a degree of freedom, as shown in figure \ref{coulombcenter}. In such a choice, all the operators associated to the boundary links commute with the rest of the operators on the algebra. Hence, they form a center.
In the $\pi$ center choice, we do it in the same way, interchanging $\hat{A}\leftrightarrow \pi$.\\
\\
\\
\\
Now let us see the results of four algebra choices. We expect the entropy to take the following form as a function of the square region size $n$,
\be
S_n=c_0+c_1n+c_{\log}\log n+\frac{c_2}{n}+\frac{c_3}{n^2}.
\ee
a) Full trivial center $-$ algebra $\mathcal{A}$\\
\begin{figure}[!h]
		\centering
		\includegraphics[width=10cm]{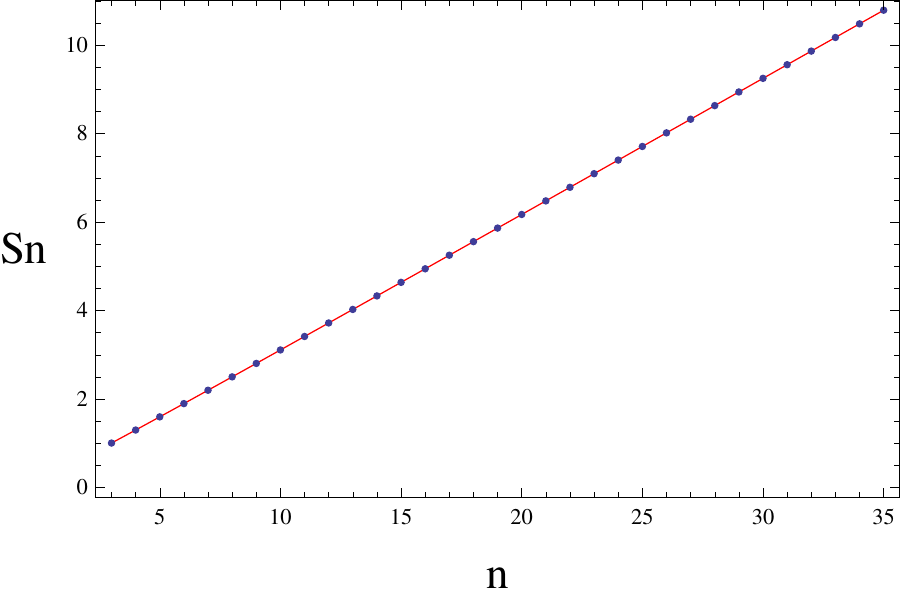}
		\caption{The entanglement entropy of gauge fields with algebra $\mathcal{A}$.}
		\label{fulltrivial}
\end{figure}\\
The entanglement entropy with algebra $\mathcal{A}$ is shown in figure \ref{fulltrivial}. We have the coefficients
\be
c_0=0.103689,\quad c_1=0.309768,\quad c_{\log}=-0.0434463,\quad c_2=0.0820914,\quad c_3=-0.0842912.
\ee
b) Trivial center with one degree of freedom removed $-$ algebra $\hat{\mathcal{A}}$\\
\begin{figure}[!h]
		\centering
		\includegraphics[width=10cm]{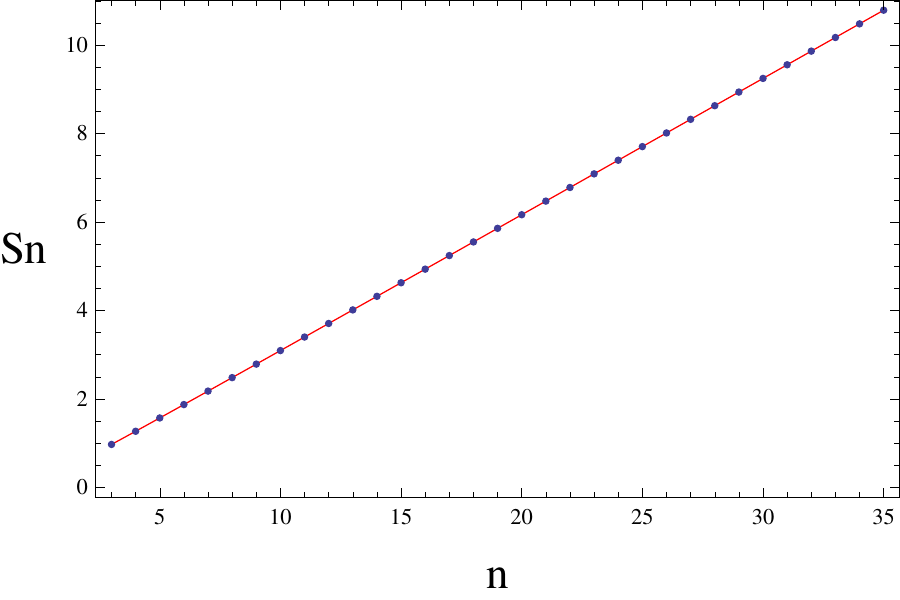}
		\caption{The entanglement entropy of gauge fields with algebra $\hat{\mathcal{A}}$.}
		\label{trivialonedofremoved}
\end{figure}\\
The entanglement entropy with algebra $\hat{\mathcal{A}}$ is shown in figure \ref{trivialonedofremoved}. We have the coefficients
\be
c_0=0.0955238,\quad c_1=0.309701,\quad c_{\log}=-0.0406814,\quad c_2=-0.0691931,\quad c_3=0.133389.
\ee
We can see that the entanglement entropy of the two different trivial centers are quite close. The effect of one degree of freedom is very small and can be neglected when the region becomes large. To calculate the entanglement entropy of gauge fields with non-trivial centers, we have to remove one physical degree of freedom after gauge fixing. The results are as follows.\\
c) $\hat{A}$ center $-$ algebra $\mathcal{A}^{\hat{A}}$\\
\begin{figure}[!h]
		\centering
		\includegraphics[width=10cm]{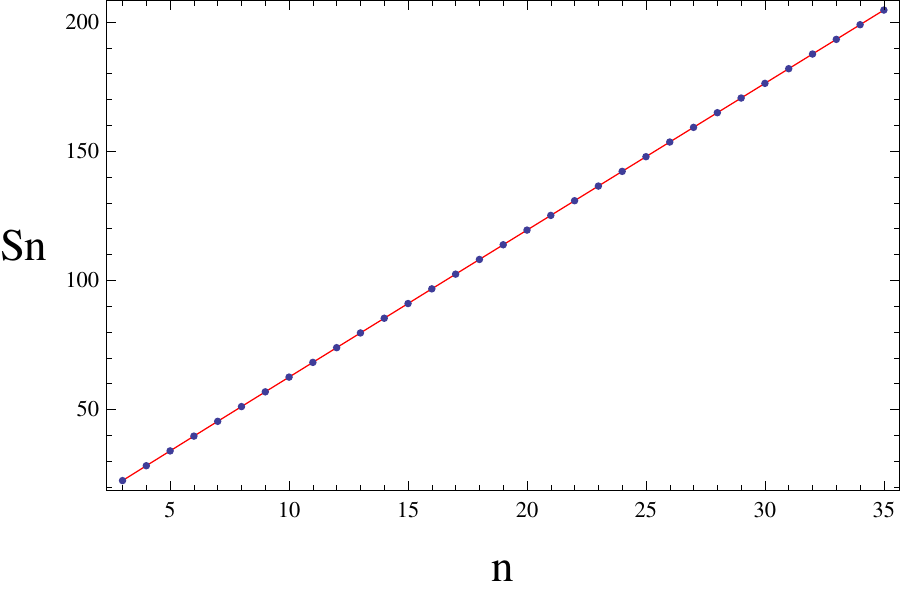}
		\caption{The entanglement entropy of gauge fields with $\mathcal{A}^{\hat{A}}$.}
		\label{coulombAcenter}
\end{figure}\\
The entanglement entropy with algebra $\mathcal{A}^{\hat{A}}$ is shown in figure \ref{coulombAcenter}. We have the coefficients
\be
c_0=4.85392,\quad c_1=5.66198,\quad c_{\log}=0.446966,\quad c_2=0.575854,\quad c_3=-0.206379.
\ee
d) $\pi$ center $-$ algebra $\mathcal{A}^{\pi}$\\
\begin{figure}[!h]
		\centering
		\includegraphics[width=10cm]{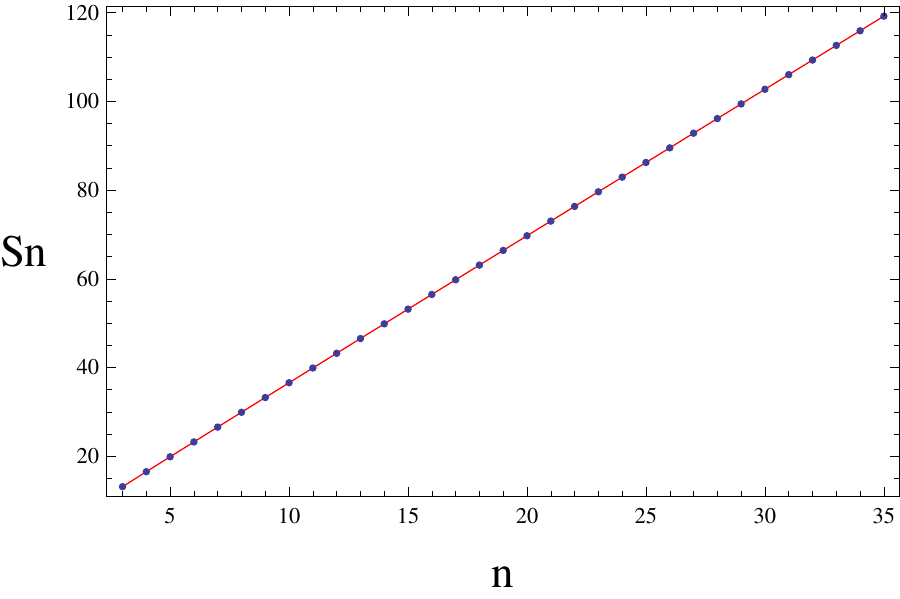}
		\caption{The entanglement entropy of gauge fields with $\mathcal{A}^{\pi}$.}
		\label{coulombPicenter}
\end{figure}\\
The entanglement entropy with algebra $\mathcal{A}^{\pi}$ is shown in figure \ref{coulombPicenter}. We have the coefficients
\be
c_0=2.59095,\quad c_1=3.2852,\quad c_{\log}=0.460412,\quad c_2=0.503424,\quad c_3=-0.0732796.
\ee
We see that, in the two trivial center choices, the logarithmic coefficients are very close, and they are also close to the logarithmic coefficient of entanglement entropy of massless scalar field in \cite{Casini:2014aia}. While, in the two non-trivial centers, the logarithmic coefficients of their entanglement entropy are also very close. They are close to the logarithmic coefficient of entanglement entropy of gauge field in \cite{Casini:2014aia}. In \cite{Casini:2014aia}, they calculate the entropy from the gauge invariant electric and magnetic fields, while we from the perspective of $\hat{A}$ and $\pi$.  The final results are very close.

\subsection{Contrasting the various center choices with the \\  electric/magnetic centers in the existing literature.}

In the previous subsection, we have studied the entanglement entropy corresponding to several different choices of algebras based on the local operators that we have constructed using the gauge potential $A_i$.
One perhaps surprising result is that by choosing a ``trivial center'' in the Coulomb gauge, it appears that the entanglement entropy agrees with that in the scalar field theory with trivial center, rather than that of the Maxwell theory \cite{Casini:2014aia}. Here we make a detailed comparison with \cite{Casini:2014aia} to explain the observation.

First of all to make meaningful comparison, we need to rewrite the operators in a common basis. Therefore the first step would be to rewrite our gauge fixed variables in terms of gauge invariant degrees of freedom.
We note that in Coulomb gauge, the magnetic field is given by $F_{12}$, where
\be
F_{ij} = \partial_i A_j - \partial_j A_i,
\ee
which implies that
\be \label{BB}
\hat{A}_j=-\nabla^2 A_j=-\partial_i F_{ij},
\ee
where the second term vanishes due to our gauge condition.
Similarly the electric field is given by
\be \label{EE}
\pi^i=\dot{A}_i=  -E_i  = \partial_0 A_i - \partial_i A_0,
\ee
where the $A_0$ field decouples entirely from our discussion as explained in the previous section. \footnote{The extra minus sign in the definition of $E$ is a convention adopted in \cite{Casini:2014aia} which we adopt for easy comparison.}

Let us first discuss the case of the trivial center in our gauge.
In terms of the scalar variable defined in (\ref{Ahatdual}) and (\ref{pidual}), and using (\ref{BB}) and (\ref{EE}), we find
\be
\pi^i=-E_i=\epsilon_{ij}\partial_j \dot{\chi}
\ee
and
\be
\hat{A}_j=-\partial_i F_{ij}=\epsilon_{jk}\partial_k \hat{\chi}.
\ee
We can see that the magnetic fields take the same place as $\hat{\chi}$.
In the lattice, from (\ref{Ahat1})$-$(\ref{pi2}), we have
\be\label{A1B}
\hat{A}_{1ij}=B_{i,j}-B_{i,j-1}=\hat{\chi}_{i,j}-\hat{\chi}_{i,j-1},
\ee
\be
\hat{A}_{2ij}=B_{i-1,j}-B_{i,j}=\hat{\chi}_{i-1,j}-\hat{\chi}_{i,j},
\ee
and
\be
\pi^1_{ij}=-E^1_{ij}=\dot{\chi}_{i,j}-\dot{\chi}_{i,j-1},
\ee
\be\label{pi2E}
\pi^2_{ij}=-E^2_{ij}=\dot{\chi}_{i-1,j}-\dot{\chi}_{i,j}.
\ee
For convenience, we take the notations
\be\label{A2scalar}
\hat{A}_{Iij}=\Delta^I_{ij}B=\Delta^I_{ij}\hat{\chi}, \qquad \pi^I_{ij}=-E^I_{ij}=\Delta^I_{ij}\dot{\chi}
\ee
where $I$ runs from $1$ to $2$, which denotes the horizontal component and vertical component respectively;  $i,j$ represent the position of the operator and $\Delta^I_{ij}$ is defined according to the equations (\ref{A1B})$-$(\ref{pi2E}).

\begin{figure}[!h]
		\centering
		\includegraphics[width=10cm]{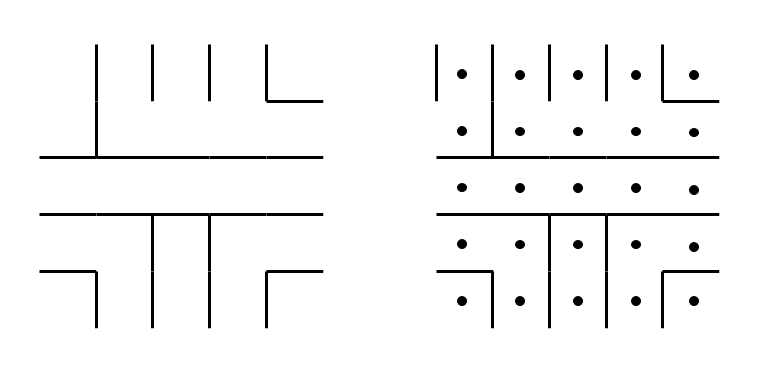}
		\caption{Contrasting  trivial centers: The left figure corresponds to the trivial center algebra of our model with square
region of size $n = 3$, while the right figure corresponds to the trivial center algebra of Casini's model \cite{Casini:2014aia} with square
region of size $n = 5$. Only physical degrees of freedom have been left in both two figures. In the left figure, links mean both the corresponding operators $\hat{A}$ and $\pi$. In the right figure, links mean the corresponding electric fields and dots means the corresponding magnetic fields. We can see that the two trivial centers are quite different. The physical operators $\hat{A}$ in the left figure can be got by truncating the magnetic fields in the right figure and then taking the maximum tree, which is a way to get the physical degrees of freedom and remove the redundancy. We will see it in figure (\ref{trivialmaximumtree}) for details.}
		\label{trivialcompare}
\end{figure}

This should be contrasted with the duality relation between the electro-magnetic fields and the scalar as presented in \cite{Casini:2014aia}, in which
\be
E^I_{ij} =\Delta^I_{ij} \phi, \qquad B = \dot \phi.
\ee
The important difference between these two sets of relations is that in our case, the fundamental degrees of freedom are the gauge operators $\hat{A}$ and $\pi$ living on the links. Therefore, the corresponding basis of operator algebra in the region is generated by $\{\pi^I_{ij}, \hat{A}_{Iij}\} \equiv\{\Delta^I_{ij} \dot \chi, \Delta^I_{ij} \hat \chi\} \equiv \{ -E^I_{ij}, \Delta^I_{ij} B\}$. On the contrary, in the usual case such as that presented in \cite{Casini:2014aia}, the operator algebra is generated by $\{E^I_{ij}, B^I_{ij}\}$. This means that even in the case of trivial center in a given region, where the number of independent $E$ and $B$ are the same, there is a global difference naturally arising as soon as we take $\hat{A}$ and $\pi$ as the basis of the operator algebra. One can check that by taking $\hat{A}$ and $\pi$ as the fundamental degrees of freedom, related to the scalars by (\ref{A2scalar}), it returns the entanglement entropy of the scalar, as opposed to the truncated scalar in \cite{Casini:2014aia}, in which only the difference of $\phi$ features in $E$ but not in $B$. This is further demonstrated in figure (\ref{trivialcompare}).
\\

\begin{figure}[!h]
		\centering
		\includegraphics[width=10cm]{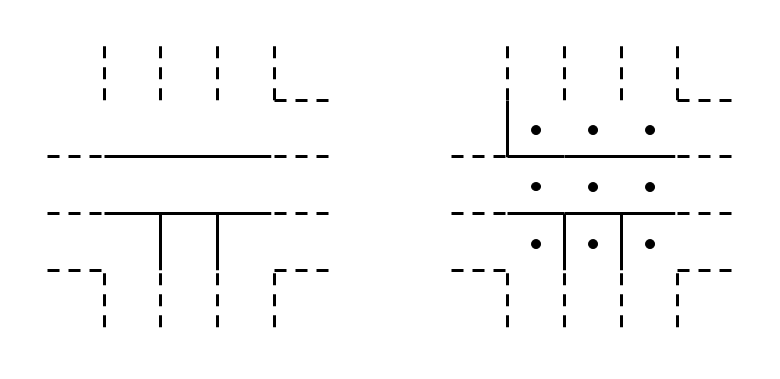}
		\caption{Comparison of non-trivial centers: The left figure corresponds to the $\hat{A}$ (or $\pi$) center algebra of our model with square
region of size $n = 3$, while the right figure corresponds to the electric center algebra of Casini's model \cite{Casini:2014aia} with square
region of size $n = 3$. Only physical degrees of freedom have been kept in both figures. In the left figure, solid links correspond to both $\hat{A}$ and $\pi$, while dashed links correspond to keeping only the operators $\hat{A}$ (or $\pi$) in the algebra, which form a center. In the right figure, both solid links and dashed links correspond to the electric operators and dots correspond to the magnetic operators. The electric operators on  dashed links commute with all other operators and form a center. We can see that the two centers (dashed links) are the same in the figures.}
		\label{centercompare}
\end{figure}

 Similar comparisons can be carried out for the case of the ``field center'' and the ``$\pi$ center''. There, one can see from figure  (\ref{centercompare}) that they coincide with the electric center taken in \cite{Casini:2014aia}.

\begin{figure}[!h]
		\centering
		\includegraphics[width=10cm]{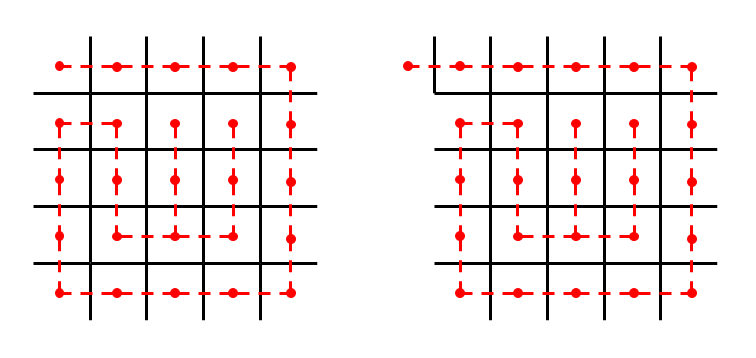}
		\caption{An example of gauge fixing: We do the gauge fixing by taking the maximal tree of dual scalars to remove the redundancy. The left figure corresponds to the gauge fixing of operators $\hat{A}$ (or $\pi$) in the trivial center algebra of our model with square region of size $n = 3$, while the right figure corresponds to the gauge fixing of electric fields in the trivial center algebra of Casini's model \cite{Casini:2014aia} with square region of size $n = 5$. In the left figure, the solid links correspond to the operators $\hat{A}$ (or $\pi$), while the red dots are the corresponding dual scalar operators $\hat{\chi}$ (or $\dot{\chi}$). The red dashed lines form a maximal tree of scalars. We remove the redundancy by removing the solid links which are not cut by red dashed lines and keep the ones which are cut by red dashed lines. In this way, we get the left figure of figure(\ref{trivialcompare}). In the right figure, the solid links mean the electric fields, while the red dots are the corresponding dual scalar fields. We do the gauge fixing in the same way. We can see that the scalar operators $\hat{\chi}$ (or $\dot{\chi}$) in the left figure take the same places as the magnetic fields in the right figure of figure (\ref{trivialcompare}). The operators $\hat{A}$ are the truncation of magnetic fields.}
		\label{trivialmaximumtree}
\end{figure}

Before we end, we would also like to make a comment about the treatment of zero mode that is adopted in \cite{Casini:2014aia} and inherited here. The entanglement entropy is computed using correlation functions of the fundamental basis of the operator algebra, $\{E, B\}$ considered in \cite{Casini:2014aia} and $\{\hat A, \pi\}$ in the current paper, expressed in terms of correlation functions of a dual scalar. In either case, there are constraints satisfied by the operators, and such redundancies are removed by picking up a maximal tree of links in the given region, which is essentially equivalent to treating the zero mode of the scalars. The notion of the maximal tree
is illustrated in figure (\ref{trivialmaximumtree}).

\section{Local Operators of $U(1)$ Gauge Fields coupling with matter}
Now we consider $U(1)$ gauge fields coupling with matter. We consider the situation of a Higgsed U(1) theory, where the gauge fields gain a mass. The Lagrangian considered is
\be
\mathcal{L}=\frac{1}{2}(\partial_\mu\sigma-m A_\mu)^2-\frac{1}{4}F_{\mu\nu}F^{\mu\nu}.
\ee
We take the gauge fixing
\be\label{fixing}
\partial_i A_i=m\sigma.
\ee
From the gauge fixing, Gauss's law reduces to
\be
\partial_i \pi^i=-\partial_i \frac{\partial\mathcal{L}}{\partial F_{i0}}=-\frac{\partial\mathcal{L}}{\partial A_0}=m(\dot{\sigma}-m A_0).
\ee
With the gauge fixing (\ref{fixing}), we have the equations of motion
\be\label{A0}
\nabla^2A_0=m^2A_0,
\ee
\be
\ddot{A}_i-\nabla^2A_i+m^2A_i=\partial_i\dot{A}_0
\ee
and
\be
\ddot{\sigma}-\nabla^2\sigma+m^2\sigma=m\dot{A}_0.
\ee
From (\ref{A0}), we find that there is no dynamic in the temporal component of gauge field. Similarly to the previous case without matter, we can take $A_0=0$ and keep the degrees of freedom of $A_i$ and $\sigma$. The canonical momenta are
\be
\pi^i=\frac{\partial \mathcal{L}}{\partial \dot{A}_i}=\dot{A}_i-\partial_i A_0=\dot{A}_i.
\ee
and
\be
\pi^\sigma=\frac{\partial \mathcal{L}}{\partial \dot{\sigma}}=\dot{\sigma}-m A_0=\dot{\sigma}.
\ee
With $A_0=0$, we get two second class constraints
\be\label{constraintA}
\chi_1=\partial_i A_i-m\sigma=0
\ee
and
\be\label{constraintpi}
\chi_2=\partial _i \pi^i-m \dot{\sigma}=0.
\ee
Because they are the second class constraints, we have to consider the Dirac bracket. The Poisson bracket is defined as
\be
\{U,V \}_P\equiv \int d^{d-1}x \left(\frac{\delta U}{\delta A_\mu(x)}\frac{\delta V}{\delta \pi^\mu(x)}-\frac{\delta V}{\delta A_\mu(x)}\frac{\delta U}{\delta \pi^\mu(x)}\right).
\ee
For the two constraints, we have
\be
\begin{aligned}
\{\chi_{1x},\chi_{2y}\}_P=&\{\partial_i A_i-m \sigma, \partial_j \pi^j-m \dot{\sigma}\}_P=\{\partial_i A_i, \partial_j \pi^j\}_P+m^2\{\sigma,\dot{\sigma}\}_P\\
=&-\nabla_x^2\delta(\vec{x}-\vec{y})+m^2\delta(\vec{x}-\vec{y})=\int \frac{d^2\vec{k}}{(2\pi)^2}(|\vec{k}|^2+m^2)e^{i\vec{k}\cdot(\vec{x}-\vec{y})}.
\end{aligned}
\ee
The Dirac bracket is defined as
\be\label{dirac bracket}
\{f,g\}_D\equiv\{f,g\}_P-\{f,\chi_i\}_P(C^{-1})_{ij}\{\chi_j,g\}_P,
\ee
where the matric $C$ is defined as
\be
C=\left(                 
  \begin{array}{cc}   
    0 & \{\chi_1,\chi_2\}_P\\  
    \{\chi_1,\chi_2\}_P & 0\\  
  \end{array}
\right).                 
\ee
For other Poisson brackets, we have
\be
\{\sigma(x),\chi_1(y)\}_P=0,
\ee
\be
\{\sigma(x),\chi_2(y)\}_P=-m\{\sigma(x),\dot{\sigma}(y)\}_P=-m\delta(\vec{x}-\vec{y}),
\ee
\be
\{\dot{\sigma}(x),\chi_1(y)\}_P=-m\{\dot{\sigma}(x),\sigma(y)\}_P=m\delta(\vec{x}-\vec{y}),
\ee
\be
\{\dot{\sigma}(x),\chi_2(y)\}_P=0,
\ee
\be
\{A_i(x),\chi_1(y)\}_P=0,
\ee
\be
\{A_i(x),\chi_2(y)\}_P=\{A_i(x),\partial_j \pi^j(y)\}_P=\partial^y_i\delta(\vec{x}-\vec{y})=-\partial_i^x\delta(\vec{x}-\vec{y}),
\ee
\be
\{\pi^i(x),\chi_1(y)\}_P=\{\pi^i(x),\partial_j A_j(y)\}_P=-\partial_i^y\delta(\vec{x}-\vec{y})=\partial_i^x\delta(\vec{x}-\vec{y}),
\ee
\be
\{\pi^i(x),\chi_2(y)\}_P=0.
\ee
From the above equations and eq(\ref{dirac bracket}), we have
\be
\{\sigma(\vec{x},t),\pi^\sigma(\vec{y},t)\}_D=\int \frac{d^2\vec{k}}{(2\pi)^2}\frac{|\vec{k}|^2}{|\vec{k}|^2+m^2}e^{i\vec{k}\dot(\vec{x}-\vec{y})},
\ee
\be
\{\sigma(\vec{x},t),\pi^i(\vec{y},t)\}_D=\int \frac{d^2\vec{k}}{(2\pi)^2}\frac{-i m k_i}{|\vec{k}|^2+m^2}e^{i\vec{k}\dot(\vec{x}-\vec{y})},
\ee
\be
\{A_i(\vec{x},t),\pi^\sigma(\vec{y},t)\}_D=\int \frac{d^2\vec{k}}{(2\pi)^2}\frac{i m k_i}{|\vec{k}|^2+m^2}e^{i\vec{k}\dot(\vec{x}-\vec{y})},
\ee
\be
\{A_i(\vec{x},t),\pi^j(\vec{y},t)\}_D=\int \frac{d^2\vec{k}}{(2\pi)^2}\left(\delta_i^j-\frac{k_i k_j}{|\vec{k}|^2+m^2}\right)e^{i\vec{k}\dot(\vec{x}-\vec{y})},
\ee
and
\be
\{\sigma(\vec{x},t),\sigma(\vec{y},t)\}_D=\{\pi^\sigma(\vec{x},t),\pi^\sigma(\vec{y},t)\}_D=\{A_i(\vec{x},t),A_j(\vec{y},t)\}_D=\{\pi^i(\vec{x},t),\pi^j(\vec{y},t)\}_D=0.
\ee
To quantize the fields, we canonically quantize, and impose the Dirac brackets to obtain the following commutators
\be
[\sigma(\vec{x},t),\pi^\sigma(\vec{y},t)]=i\int \frac{d^2\vec{k}}{(2\pi)^2}\frac{|\vec{k}|^2}{|\vec{k}|^2+m^2}e^{i\vec{k}\dot(\vec{x}-\vec{y})},
\ee
\be
[\sigma(\vec{x},t),\pi^i(\vec{y},t)]=i\int \frac{d^2\vec{k}}{(2\pi)^2}\frac{-i m k_i}{|\vec{k}|^2+m^2}e^{i\vec{k}\dot(\vec{x}-\vec{y})},
\ee
\be
[A_i(\vec{x},t),\pi^\sigma(\vec{y},t)]=i\int \frac{d^2\vec{k}}{(2\pi)^2}\frac{i m k_i}{|\vec{k}|^2+m^2}e^{i\vec{k}\dot(\vec{x}-\vec{y})},
\ee
\be
[A_i(\vec{x},t),\pi^j(\vec{y},t)]=i\int \frac{d^2\vec{k}}{(2\pi)^2}\left(\delta_i^j-\frac{k_i k_j}{|\vec{k}|^2+m^2}\right)e^{i\vec{k}\dot(\vec{x}-\vec{y})},
\ee
and
\be
[\sigma(\vec{x},t),\sigma(\vec{y},t)]=[\pi^\sigma(\vec{x},t),\pi^\sigma(\vec{y},t)]=[A_i(\vec{x},t),A_j(\vec{y},t)]=[\pi^i(\vec{x},t),\pi^j(\vec{y},t)]=0.
\ee
Here we know, there are 2 degrees of physical freedom in the total fields in 2+1 dimension. For simplicity, we treat $\sigma$ as $A_0$. We do the mode expansion and we get
\be
A_\mu(\vec{x},t)=\int \frac{d^2\vec{k}}{2\pi}\frac{1}{\sqrt{2\omega}}\sum_{\sigma}\left(e_\mu(\vec{k},\sigma)\hat{a}(\vec{k},\sigma)e^{-i\omega t+i\vec{k}\cdot\vec{x}}+e_\mu(\vec{k},\sigma)^*\hat{a}^\dagger(\vec{k},\sigma)e^{i\omega t-i\vec{k}\cdot\vec{x}}\right)
\ee
and
\be
\pi^\mu(\vec{x},t)=\int \frac{d^2\vec{k}}{2\pi}\frac{1}{\sqrt{2\omega}}\sum_{\sigma}\left(-i\omega e_\mu(\vec{k},\sigma)\hat{a}(\vec{k},\sigma)e^{-i\omega t+i\vec{k}\cdot\vec{x}}+i \omega e_\mu(\vec{k},\sigma)^*\hat{a}^\dagger(\vec{k},\sigma)e^{i\omega t-i\vec{k}\cdot\vec{x}}\right).
\ee
Here $\omega^2=|\vec{k}|^2+m^2$. From the above Dirac brackets, we have the constraints for the polarizations
\be\label{e0e0}
\sum_{\sigma}\left(e_0(\vec{k},\sigma)e_0(\vec{k},\sigma)^*\right)=\frac{|\vec{k}|^2}{|\vec{k}|^2+m^2},
\ee
\be\label{e0ei}
\sum_{\sigma}\left(e_0(\vec{k},\sigma)e_i(\vec{k},\sigma)^*\right)=-\frac{i m k_i}{|\vec{k}|^2+m^2},
\ee
\be\label{eie0}
\sum_{\sigma}\left(e_i(\vec{k},\sigma)e_0(\vec{k},\sigma)^*\right)=\frac{i m k_i}{|\vec{k}|^2+m^2},
\ee
\be\label{eiej}
\sum_{\sigma}\left(e_i(\vec{k},\sigma)e_j(\vec{k},\sigma)^*\right)=\delta_{ij}-\frac{k_i k_j}{|\vec{k}|^2+m^2}.
\ee
However, from the commutators above, we find that the operators $\sigma$, $A_i$ and their canonical momentums are again not local. As before, we need to construct a set of local operators in this model. We define
\be
\tilde{A}_i\equiv-\nabla^2(-\nabla^2+m^2)A_i,
\ee
\be
\tilde{\sigma}\equiv-\nabla^2(-\nabla^2+m^2)\sigma,
\ee
and consider the operators $\tilde{\sigma}$,$\tilde{A}_i$ and $\pi^\sigma$,$\pi^j$. The two constraints of the new variables are
\be\label{Atildefixing}
\partial_i \tilde{A}_i=m\tilde{\sigma}
\ee
and
\be\label{pitildefixing}
\partial _i \pi^i-m \dot{\sigma}=0.
\ee
The commutators of the new operators are
\be
[\tilde{\sigma}(\vec{x},t),\pi^\sigma(\vec{y},t)]=i\int \frac{d^2\vec{k}}{(2\pi)^2}|\vec{k}|^4e^{i\vec{k}\dot(\vec{x}-\vec{y})},
\ee
\be
[\tilde{\sigma}(\vec{x},t),\pi^i(\vec{y},t)]=i\int \frac{d^2\vec{k}}{(2\pi)^2}(-i|\vec{k}|^2 m k_i)e^{i\vec{k}\dot(\vec{x}-\vec{y})},
\ee
\be
[\tilde{A}_i(\vec{x},t),\pi^\sigma(\vec{y},t)]=i\int \frac{d^2\vec{k}}{(2\pi)^2}i|\vec{k}|^2 m k_ie^{i\vec{k}\dot(\vec{x}-\vec{y})},
\ee
\be
[\tilde{A}_i(\vec{x},t),\pi^j(\vec{y},t)]=i\int \frac{d^2\vec{k}}{(2\pi)^2}\left(\delta_i^j|\vec{k}|^2 (|\vec{k}|^2+m^2)-k_i k_j |\vec{k}|^2 \right)e^{i\vec{k}\dot(\vec{x}-\vec{y})},
\ee
and
\be
[\tilde{\sigma}(\vec{x},t),\tilde{\sigma}(\vec{y},t)]=[\pi^\sigma(\vec{x},t),\pi^\sigma(\vec{y},t)]=[\tilde{A}_i(\vec{x},t),\tilde{A}_j(\vec{y},t)]=[\pi^i(\vec{x},t),\pi^j(\vec{y},t)]=0.
\ee
We can see that the new operators are local. We will discretize them in the lattice and calculate the entanglement entropy. In the next section, we will find that the new operators are very useful when we consider  the lattice.

\section{$U(1)$ Gauge Fields coupling with matter on the lattice}
For $U(1)$ gauge fields coupling with matter $\sigma$, we have to modify the duality relation (\ref{eq:dual1}).
\be
A_i=\partial_i\phi+\epsilon_{ij}\partial_j\chi,
\ee
\be
m\sigma=\nabla^2\phi,
\ee
and
\be
\pi^i=\partial_i\dot{\phi}+\epsilon_{ij}\partial_j\dot{\chi},
\ee
\be
m\pi^\sigma=\nabla^2\dot{\phi},
\ee
which satisfies the constraints (\ref{constraintA}) and (\ref{constraintpi}) automatically. We have the mode expansions of the dual scalar fields and the time derivative of them
\be
\begin{aligned}
\chi(\vec{x},t)=-\int\frac{d^2\vec{k}}{2\pi}\frac{1}{\sqrt{2\omega}}\frac{1}{|\vec{k}|^2}\sum_{\sigma}&\left((i k_y e_1(\vec{k},\sigma)-i k_x e_2(\vec{k},\sigma))\hat{a}(\vec{k},\sigma)e^{-i\omega t+i\vec{k}\cdot\vec{x}}\right.\\
&\left.-(i k_y e_1(\vec{k},\sigma)^*-i k_x e_2(\vec{k},\sigma)^*)\hat{a}^\dagger(\vec{k},\sigma)e^{i\omega t-i\vec{k}\cdot\vec{x}}\right),
\end{aligned}
\ee
\be
\phi(\vec{x},t)=-m\int\frac{d^2\vec{k}}{2\pi}\frac{1}{\sqrt{2\omega}}\sum_{\sigma}\frac{1}{|\vec{k}|^2}\left(e_0(\vec{k},\sigma)\hat{a}(\vec{k},\sigma)e^{-i\omega t+i\vec{k}\cdot\vec{x}}+e_0(\vec{k},\sigma)^*\hat{a}^\dagger(\vec{k},\sigma)e^{i\omega t-i\vec{k}\cdot\vec{x}}\right),
\ee
and
\be
\begin{aligned}
\dot{\chi}(\vec{x},t)=-\int\frac{d^2\vec{k}}{2\pi}\sqrt{\frac{\omega}{2}}\frac{1}{|\vec{k}|^2}\sum_{\sigma}&\left((k_y e_1(\vec{k},\sigma)-k_x e_2(\vec{k},\sigma))\hat{a}(\vec{k},\sigma)e^{-i\omega t+i\vec{k}\cdot\vec{x}}\right.\\
&\left.+(k_y e_1(\vec{k},\sigma)^*-k_x e_2(\vec{k},\sigma)^*)\hat{a}^\dagger(\vec{k},\sigma)e^{i\omega t-i\vec{k}\cdot\vec{x}}\right),
\end{aligned}
\ee
\be
\dot{\phi}(\vec{x},t)=-m\int\frac{d^2\vec{k}}{2\pi}\sqrt{\frac{\omega}{2}}\sum_{\sigma}\left(-i e_0(\vec{k},\sigma)\hat{a}(\vec{k},\sigma)e^{-i\omega t+i\vec{k}\cdot\vec{x}}+i e_0(\vec{k},\sigma)^*\hat{a}^\dagger(\vec{k},\sigma)e^{i\omega t-i\vec{k}\cdot\vec{x}}\right).
\ee
Similarly, for the new operators $\tilde{\sigma}$,$\tilde{A}_i$ and $\pi^\sigma$,$\pi^j$, we have the duality relation
\be\label{Atildedual}
\tilde{A}_i=\partial_i\tilde{\phi}+\epsilon_{ij}\partial_j\tilde{\chi},
\ee
\be
m\tilde{\sigma}=\nabla^2\tilde{\phi},
\ee
\be\label{pitildedual}
\pi^i=\partial_i\dot{\phi}+\epsilon_{ij}\partial_j\dot{\chi},
\ee
\be
m\pi^\sigma=\nabla^2\dot{\phi},
\ee
where we define
\be
\begin{aligned}
\tilde{\chi}(\vec{x},t)&\equiv -\nabla^2(-\nabla^2+m^2)\phi\\
&=-\int\frac{d^2\vec{k}}{2\pi}\frac{1}{\sqrt{2\omega}}\omega^2
\sum_{\sigma}\left((i k_y e_1(\vec{k},\sigma)-i k_x e_2(\vec{k},\sigma))\hat{a}(\vec{k},\sigma)e^{-i\omega t+i\vec{k}\cdot\vec{x}}\right.\\
&\left.-(i k_y e_1(\vec{k},\sigma)^*-i k_x e_2(\vec{k},\sigma)^*)\hat{a}^\dagger(\vec{k},\sigma)e^{i\omega t-i\vec{k}\cdot\vec{x}}\right),
\end{aligned}
\ee
and
\be
\begin{aligned}
\tilde{\phi}(\vec{x},t)&\equiv -\nabla^2(-\nabla^2+m^2)\phi\\
&=-m\int\frac{d^2\vec{k}}{2\pi}\frac{1}{\sqrt{2\omega}}\omega^2\sum_{\sigma}\left(e_0(\vec{k},\sigma)\hat{a}(\vec{k},\sigma)e^{-i\omega t+i\vec{k}\cdot\vec{x}}+e_0(\vec{k},\sigma)^*\hat{a}^\dagger(\vec{k},\sigma)e^{i\omega t-i\vec{k}\cdot\vec{x}}\right).
\end{aligned}
\ee
\begin{figure}[!h]
		\centering
		\includegraphics[width=8cm]{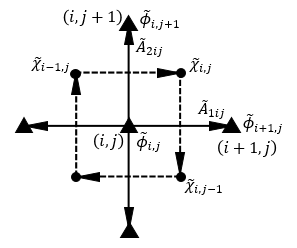}
		\caption{Dual lattice: The scalar field operator $\tilde{\chi}$ is in the center of the plaquette and operator $\tilde{\phi}$ is in the vertex. The gauge field operator $\tilde{A}$ in some link is equal to a difference of scalar field operators $\tilde{\chi}$ across the link in the dual lattice which is perpendicular to the one corresponding to $\tilde{A}$ plus a difference of scalar field operators $\tilde{\phi}$ along the same link in the dual lattice. The duality of gauge field momentum operator $\pi$ is in the same way.}
		\label{matterdual}
\end{figure}\\
The non-vanishing commutators of $\tilde{\chi}$,$\dot{\chi}$,$\tilde{\phi}$ and $\dot{\phi}$ are
\be
[\tilde{\chi}(\vec{x},t),\dot{\chi}(\vec{y},t)]=i\int\frac{d^2\vec{k}}{(2\pi)^2}\omega^2e^{i\vec{k}\cdot(\vec{x}-\vec{y})}
\ee
and
\be
[\tilde{\phi}(\vec{x},t),\dot{\phi}(\vec{y},t)]=im^2\delta^2(\vec{x}-\vec{y}).
\ee
The non-vanishing vacuum correlation functions are
\be
\langle\tilde{\chi}(\vec{x},t)\tilde{\chi}(\vec{y},t)\rangle=\int\frac{d^2\vec{k}}{(2\pi)^2}\frac{\omega^3}{2}|\vec{k}|^2e^{i\vec{k}\cdot(\vec{x}-\vec{y})},
\ee
\be
\langle\dot{\chi}(\vec{x},t)\dot{\chi}(\vec{y},t)\rangle=\int\frac{d^2\vec{k}}{(2\pi)^2}\frac{\omega}{2}\frac{1}{|\vec{k}|^2}e^{i\vec{k}\cdot(\vec{x}-\vec{y})},
\ee
\be
\langle\tilde{\phi}(\vec{x},t)\tilde{\phi}(\vec{y},t)\rangle=m^2\int\frac{d^2\vec{k}}{(2\pi)^2}\frac{\omega}{2}|\vec{k}|^2e^{i\vec{k}\cdot(\vec{x}-\vec{y})},
\ee
and
\be
\langle\dot{\phi}(\vec{x},t)\dot{\phi}(\vec{y},t)\rangle=m^2\int\frac{d^2\vec{k}}{(2\pi)^2}\frac{\omega}{2}\frac{1}{|\vec{k}|^2\omega^2}e^{i\vec{k}\cdot(\vec{x}-\vec{y})}.
\ee
Now we discretize the model in a square lattice. We define the operators $\tilde{A}_1$ and $\pi^1$ associated to horizontal links, $\tilde{A}_2$ and $\pi^2$ to vertical links, $\tilde{\sigma}$ and $\pi^\sigma$ to the vertices. Because of the redundant degrees of freedom in this model, we have two constraints (\ref{Atildefixing}) and (\ref{pitildefixing}). In the discrete lattice, the two constraints become
\be
\sum_b \tilde{A}_{ab}=m\tilde{\sigma}_a
\ee
and
\be
\sum_b \tilde{\pi}^{ab}=m\pi^{\sigma a},
\ee
where the sum is over all the links $(ab)$ with the common vertex $a$. In the above equations, it is assumed that the gauge field component is the corresponding one to the link direction and matter field component is associated to the vertex. The links are oriented. The field attached to changes sign if the orientation is flipped i.e. $\tilde{A}_{ab}=-\tilde{A}_{ba}$.\\

Because of the above constraints, we fix the scalar field variables $\tilde{\sigma}$ and $\pi^{\sigma}$ to make gauge fields the physical degrees of freedom. As shown in figure (\ref{matterdual}), we don't show the scalar field variable $\tilde{\sigma}$ there because we have fixed it. The way to label the gauge field operators on the lattice is the same as the pure gauge theory in section 4. We label them with the coordinates of the initial vertices of the vectors, such as
\be
\tilde{A}_{1ij}\equiv \tilde{A}_{1(ij,i+1j)},
\ee
\be
\tilde{A}_{2ij}\equiv \tilde{A}_{2(ij,ij+1)}.
\ee
The discrete version of (\ref{Atildedual}) and (\ref{pitildedual}) is also shown in figure (\ref{matterdual}). The operator $\tilde{A}$ is related to the difference of the scalar field operators $\tilde{\chi}$ in the orthogonal direction in the dual lattice and the difference of the scalar field operators $\tilde{\phi}$ at the two end vertices of the same link as $\tilde{A}$. The dual of operator $\pi$ is defined in the same way, interchanging $\tilde{A} \leftrightarrow \pi$, $\tilde{\chi} \leftrightarrow \dot{chi}$ and $\tilde{\phi} \leftrightarrow \dot{\phi}$. For example,
\be
\tilde{A}_{1ij}=\tilde{\chi}_{i,j}-\tilde{\chi}_{i,j-1}+\tilde{\phi}_{i+1,j}-\tilde{\phi}_{i,j},
\ee
\be
\tilde{A}_{2ij}=\tilde{\chi}_{i-1,j}-\tilde{\chi}_{i,j}+\tilde{\phi}_{i,j+1}-\tilde{\phi}_{i,j},
\ee
and
\be
\pi^1_{ij}=\dot{\chi}_{i,j}-\dot{\chi}_{i,j-1}+\dot{\phi}_{i+1,j}-\dot{\phi}_{i,j},
\ee
\be
\pi^2_{ij}=\dot{\chi}_{i-1,j}-\dot{\chi}_{i,j}+\dot{\phi}_{i,j+1}-\dot{\phi}_{i,j}.
\ee
With the above dualities, the non-vanishing commutators of the discrete version of operators $\tilde{\chi}$, $\dot{\chi}$ and $\tilde{\phi}$, $\dot{\phi}$ are
\be \label{eq:commutechi}
[\tilde{\chi}_{ij},\dot{\chi}_{kl}]=im^2\delta_{ik}\delta_{jl}+\frac{i}{(2\pi)^2}\int_{-\pi}^{\pi}d k_x\int_{-\pi}^{\pi}d k_y(4\sin^2\frac{k_x}{2}+4\sin^2\frac{k_y}{2})\cos(k_x(i-k))\cos(k_y(j-l)),
\ee
and
\be
[\tilde{\phi}_{ij},\dot{\phi}_{kl}]=im^2\delta_{ik}\delta_{jl}.
\ee
Note that in the second term on the rhs of (\ref{eq:commutechi}), we have left the result inside an integral. It might at first sight lead one to question the locality of $\tilde\chi$. We note however that these terms are precisely corresponding to the derivatives of the delta function. They are thus local on the lattice. We leave them in the current form for simpler manipulation in the numerics.
The commutators of the discrete version of operators $\tilde{A}$, $\pi$ can be expressed by the above commutators. The non-vanishing ones are
\be
\begin{aligned}
\left[\tilde{A}_{1ij},\tilde{A}_{1kl}\right]= & 2[\tilde{\chi}_{i,j},\tilde{\chi}_{k,l}]-[\tilde{\chi}_{i,j},\tilde{\chi}_{k,l-1}]-[\tilde{\chi}_{i,j-1},\tilde{\chi}_{k,l}]\\
& +2[\tilde{\phi}_{i,j},\tilde{\phi}_{k,l}]-[\tilde{\phi}_{i+1,j},\tilde{\phi}_{k,l}]-[\tilde{\phi}_{i,j},\tilde{\phi}_{k+1,l}],
\end{aligned}
\ee
\be
\begin{aligned}
\left[\tilde{A}_{1ij},\tilde{A}_{2kl}\right]=&[\tilde{\chi}_{i,j},\tilde{\chi}_{k-1,l}]+[\tilde{\chi}_{i,j-1},\tilde{\chi}_{k,l}]-[\tilde{\chi}_{i,j},\tilde{\chi}_{k,l}]-[\tilde{\chi}_{i,j-1},\tilde{\chi}_{k-1,l}]\\
&+[\tilde{\phi}_{i+1,j},\tilde{\phi}_{k,l+1}]+[\tilde{\phi}_{i,j},\tilde{\phi}_{k,l}]-[\tilde{\phi}_{i+1,j},\tilde{\phi}_{k,l}]-[\tilde{\phi}_{i,j},\tilde{\phi}_{k,l+1}],
\end{aligned}
\ee
\be
\begin{aligned}
\left[\tilde{A}_{2ij},\tilde{A}_{1kl}\right]=&[\tilde{\chi}_{i-1,j},\tilde{\chi}_{k,l}]+[\tilde{\chi}_{i,j},\tilde{\chi}_{k,l-1}]-[\tilde{\chi}_{i,j},\tilde{\chi}_{k,l}]-[\tilde{\chi}_{i-1,j},\tilde{\chi}_{k,l-1}]\\
&+[\tilde{\phi}_{i,j+1},\tilde{\phi}_{k+1,l}]+[\tilde{\phi}_{i,j},\tilde{\phi}_{k,l}]-[\tilde{\phi}_{i,j+1},\tilde{\phi}_{k,l}]-[\tilde{\phi}_{i,j},\tilde{\phi}_{k+1,l}],
\end{aligned}
\ee
and
\be
\begin{aligned}
\left[\tilde{A}_{1ij},\tilde{A}_{1kl}\right]=&2[\tilde{\chi}_{i,j},\tilde{\chi}_{k,l}]-[\tilde{\chi}_{i-1,j},\tilde{\chi}_{k,l}]-[\tilde{\chi}_{i,j},\tilde{\chi}_{k-1,l}]\\
&+2[\tilde{\phi}_{i,j},\tilde{\phi}_{k,l}]-[\tilde{\phi}_{i,j+1},\tilde{\phi}_{k,l}]-[\tilde{\phi}_{i,j},\tilde{\phi}_{k,l+1}].
\end{aligned}
\ee
We can see that the discrete version of operators $\tilde{A}$ and $\pi$ are local. We use them to calculate the entanglement entropy in section 8.\\
The non-vanishing vacuum correlation functions of operators $\tilde{\chi}$, $\dot{\chi}$ and $\tilde{\phi}$, $\dot{\phi}$ are
\be
\begin{aligned}
\langle\tilde{\chi}_{ij}\tilde{\chi}_{kl}\rangle=&\frac{1}{(2\pi)^2}\int_{-\pi}^{\pi}d k_x\int_{-\pi}^{\pi}d k_y\frac{1}{2}\left(4\sin^2\frac{k_x}{2}+4\sin^2\frac{k_y}{2}+m^2\right)^{\frac{3}{2}}\\
&(4\sin^2\frac{k_x}{2}+4\sin^2\frac{k_y}{2})\cos(k_x(i-k))\cos(k_y(j-l)),
\end{aligned}
\ee
\be
\langle\dot{\chi}_{ij}\dot{\chi}_{kl}\rangle=\frac{1}{(2\pi)^2}\int_{-\pi}^{\pi}d k_x\int_{-\pi}^{\pi}d k_y\frac{1}{2}\sqrt{4\sin^2\frac{k_x}{2}+4\sin^2\frac{k_y}{2}+m^2}\frac{\cos(k_x(i-k))\cos(k_y(j-l))}{4\sin^2\frac{k_x}{2}+4\sin^2\frac{k_y}{2}},
\ee
\be
\begin{aligned}
\langle\tilde{\phi}_{ij}\tilde{\phi}_{kl}\rangle=&\frac{1}{(2\pi)^2}\int_{-\pi}^{\pi}d k_x\int_{-\pi}^{\pi}d k_y\frac{m^2}{2}\sqrt{(4\sin^2\frac{k_x}{2}+4\sin^2\frac{k_y}{2}+m^2)}\\
&(4\sin^2\frac{k_x}{2}+4\sin^2\frac{k_y}{2})\cos(k_x(i-k))\cos(k_y(j-l)),
\end{aligned}
\ee
and
\be
\langle\dot{\phi}_{ij}\dot{\phi}_{kl}\rangle=\frac{1}{(2\pi)^2}\int_{-\pi}^{\pi}d k_x\int_{-\pi}^{\pi}d k_y\frac{m^2}{2\sqrt{4\sin^2\frac{k_x}{2}+4\sin^2\frac{k_y}{2}+m^2}}\frac{\cos(k_x(i-k))\cos(k_y(j-l))}{4\sin^2\frac{k_x}{2}+4\sin^2\frac{k_y}{2}}.
\ee
The vacuum correlation functions of discrete variables $\tilde{A}_{ij}$ and $\pi_{ij}$ can be expressed with the above correlation functions, such as
\be
\begin{aligned}
\langle\tilde{A}_{1ij},\tilde{A}_{1kl}\rangle=&2\langle\tilde{\chi}_{i,j},\tilde{\chi}_{k,l}\rangle-\langle\tilde{\chi}_{i,j},\tilde{\chi}_{k,l-1}\rangle-\langle\tilde{\chi}_{i,j-1},\tilde{\chi}_{k,l}\rangle\\
&+2\langle\tilde{\phi}_{i,j},\tilde{\phi}_{k,l}\rangle-\langle\tilde{\phi}_{i+1,j},\tilde{\phi}_{k,l}\rangle-\langle\tilde{\phi}_{i,j},\tilde{\phi}_{k+1,l}\rangle.
\end{aligned}
\ee

\section{Entanglement Entropy of two dimensional lattice gauge fields in the Higgs phase}
\begin{figure}[!h]
		\centering
		\includegraphics[width=12cm]{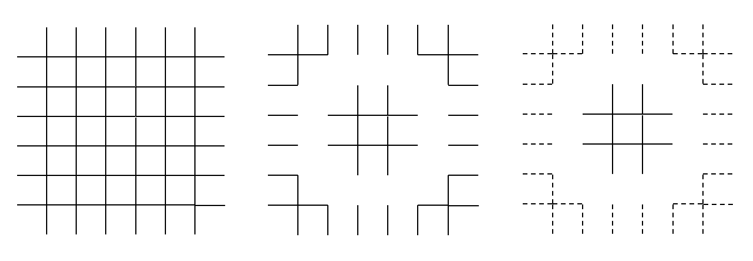}
		\caption{Some algebra choices of gauge field with matter. The figures correspond to square regions of size $n=5$. By fixing all the matter field operators on the vertices, all operators on the links are physical. In the left figure, we keep all the physical degrees of freedom and get the algebra of trivial center. We denote it as $\mathcal{B}$. In the middle figure, we remove some links. In the remaining links, both operators $\tilde{A}$ and $\pi$ are there. It also forms an algebra of trivial center. We denote it as $\tilde{\mathcal{B}}$. In the right figure, links with solid lines mean both the corresponding operators $\tilde{A}$ and $\pi$ belong to the algebra and links with dashed lines mean the corresponding operator $\tilde{A}$ or$\pi$ does not belong to the algebra. Without loss of generality, let us assume operators $\tilde{A}$ to be on the dashed links, not $\pi$. The operators $\tilde{A}$ on dashed links commute with all the operators $\pi$ on the solid links and they form a center. We denote such an algebra as $\mathcal{B}^{\tilde{A}}$. When the dashed links correspond to operators $\pi$, not $\tilde{A}$, we get the algebra of $\pi$ center. we denote it as $\mathcal{B}^{\pi}$.}
		\label{matterphysical}
\end{figure}

We calculate the entanglement entropy of gauge fields coupling with matter in a square region now. We consider the square region with four different algebra choices, which are shown in figure (\ref{matterphysical}). The four choices of algebras are full trivial center $\mathcal{B}$, trivial center with some physical degrees of freedom removed $\tilde{\mathcal{B}}$, $\tilde{A}$ center $\mathcal{B}^{\tilde{A}}$ and $\pi$ center $\mathcal{B}^{\pi}$.

\begin{figure}[!h]
		\centering
		\includegraphics[width=12cm]{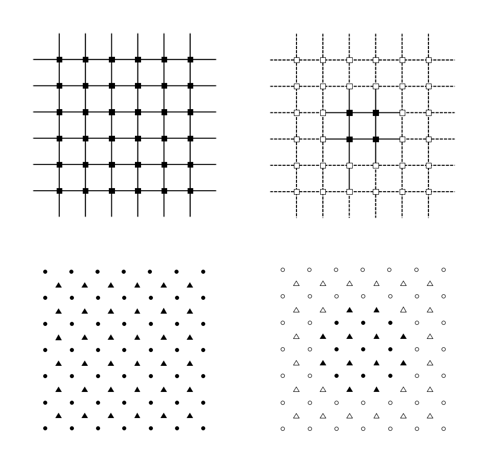}
		\caption{The duality of gauge fields with matter on the lattice. The figures are corresponding to square regions of size $n=5$. The top two figures correspond to the gauge field and matter field, while the bottom ones to the dual scalar fields representation of the same algebra. Links with solid lines mean both the  operators $\tilde{A}$ and $\pi$ on the link belong to the algebra. Links with dashed lines mean the corresponding operator $\tilde{A}$ or $\pi$ does not belong to the algebra. Marked boxes correspond to both the matter field operators $\tilde{\sigma}$ and $\pi^\sigma$. Marked dots correspond to both the scalar field operators $\tilde{\chi}$ and the momentum operator $\dot{\chi}$. Circle dots mean the scalar field operator $\tilde{\chi}$ or operator $\dot{\chi}$ does not belong to the algebra. Marked triangles correspond to both the scalar field operators $\tilde{\phi}$ and the momentum operator $\dot{\phi}$. Unmarked triangles mean the scalar field operator $\tilde{\phi}$ or the momentum operator $\dot{\phi}$ does not belong to the algebra. Because of the constraints of gauge field and matter field, we fix the matter field to remove the redundant degrees of freedom. The gauge field operators correspond to the physical degrees of freedom. The left panel shows the trivial center choice, while the right panel is related to the $\tilde{A}$ center or $\pi$ center choice (not exactly), according to the meaning of dashed lines.}
		\label{latticematter}
\end{figure}

In figure (\ref{latticematter}), we illustrate the different algebras in the different duality frames. In the top figures, we show the gauge field operators and matter field operators on the lattice, while in the bottom figures, we show the corresponding dual scalar field operators. The figure (\ref{latticematter}) is without gauge fixing. We remove the redundant degrees of freedom by fixing the variables $\tilde{\sigma}$ and $\pi^\sigma$. By gauge fixing, we get the full trivial center algebra from the left panel of figure (\ref{latticematter}). After fixing the matter degrees of freedom in the right panel of figure (\ref{latticematter}) and remove some operators, we can get the $\tilde{A}$ center algebra or the$\pi$ center algebra, as shown in figure (\ref{matterphysical}).\\

In the trivial center choices, as shown in the first and second figure of figure (\ref{matterphysical}), both operators $\tilde{A}$ and $\pi$ are associated to every link. We have the same number of operators $\tilde{A}$ and $\pi$. In the first figure, we keep all the physical degrees of freedom, while in the second figure, we remove some links to get another algebra.\\

To get the $\tilde{A}$ center choice, we remove operators $\pi$ on the boundary links of the second figure in figure (\ref{matterphysical}) which leads to the third figure. We find that all the operators $\tilde{A}$ associated to the boundary links commute with the rest of the operators in the algebra. Hence, they form a center. In the $\pi$ center choice, we do it in the same way, interchanging $\tilde{A}\leftrightarrow \pi$.\\

\begin{figure}[!h]
		\centering
		\includegraphics[width=10cm]{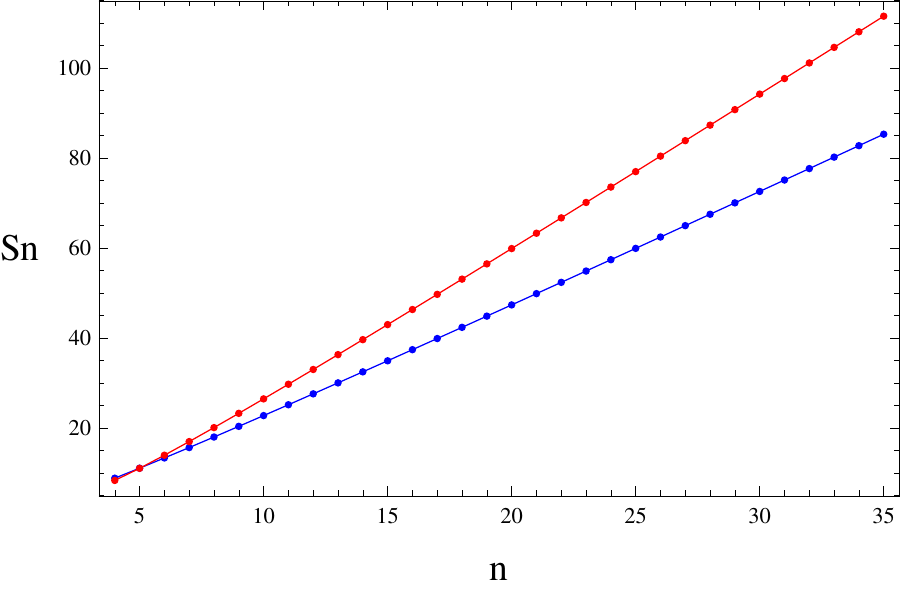}
		\caption{The entanglement entropy of gauge field with two different trivial centers.The red (top) line corresponds to the trivial with degree of freedoms removed, while the blue (bottom) line corresponds to the usual trivial center.}
		\label{mattertwotrivialcenters}
\end{figure}

Now let us see the results of fours algebra choices. We also expect the entropy has the following form as a function of the square region size $n$,
\be
S_n=c_0+c_1n+c_{\log}\log n+\frac{c_2}{n}+\frac{c_3}{n^2}.
\ee
a) Full trivial center $-$ algebra $\mathcal{B}$\\
\begin{figure}[!h]
		\centering
		\includegraphics[width=10cm]{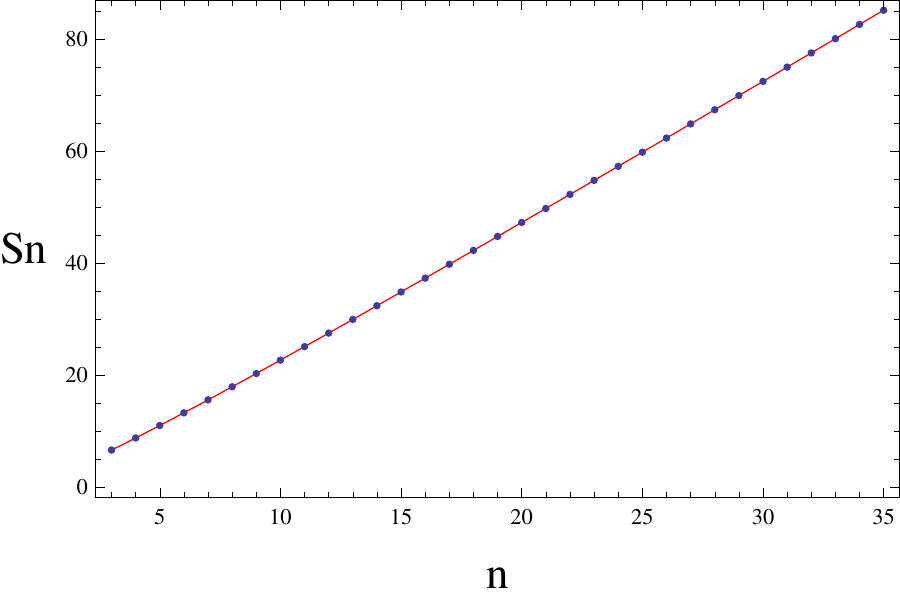}
		\caption{The entanglement entropy of gauge field with algebra $\mathcal{B}$.}
		\label{mattertrivialcenter}
\end{figure}\\
The entanglement entropy with algebra $\mathcal{B}$ is shown in figure (\ref{mattertrivialcenter}). We have the coefficients
\be
c_0=3.07418,\quad c_1=2.61987,\quad c_{\log}=-2.6416,\quad c_2=-4.77179,\quad c_3=2.12216.
\ee
b) Trivial center with some physical degrees of freedom removed $-$ algebra $\tilde{\mathcal{B}}$\\
\begin{figure}[!h]
		\centering
		\includegraphics[width=10cm]{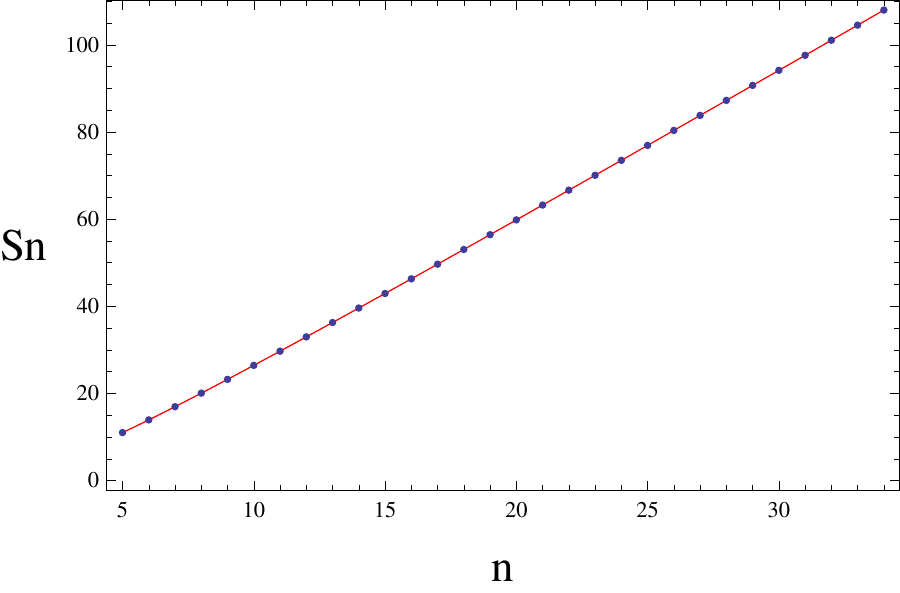}
		\caption{The entanglement entropy of gauge field with trivial center with degree of freedoms removed.}
		\label{mattertrivialcenterdofremoved}
\end{figure}\\
The entanglement entropy with algebra $\tilde{\mathcal{B}}$ is shown in figure (\ref{mattertrivialcenterdofremoved}). We have the coefficients
\be
c_0=-3.3877,\quad c_1=3.54833,\quad c_{\log}=-2.66965,\quad c_2=5.20503,\quad c_3=-1.97591.
\ee
We show the entanglement entropy with two different trivial centers together in figure (\ref{mattertwotrivialcenters}). We can see that the leading term of entanglement entropy is different for two trivial centers, but from the fitting coefficients, we find that the logarithmic term is very close.
c) $\tilde{A}$ center $-$ algebra $\mathcal{B}^{\tilde{A}}$\\
\begin{figure}[!h]
		\centering
		\includegraphics[width=10cm]{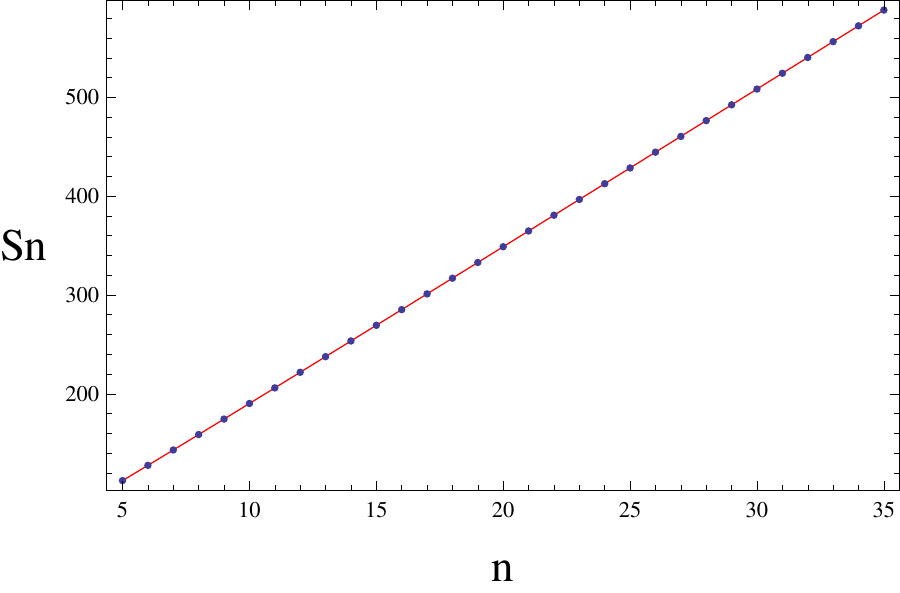}
		\caption{The entanglement entropy of gauge field with $\tilde{A}$ center.}
		\label{matterAcenter}
\end{figure}\\
The entanglement entropy with algebra $\mathcal{B}^{\tilde{A}}$ is shown in figure (\ref{matterAcenter}). We have the coefficients
\be
c_0=35.1262,\quad c_1=16.0801,\quad c_{\log}=-2.70508,\quad c_2=4.65304,\quad c_3=4.65304.
\ee
d) $\pi$ center $-$ algebra $\mathcal{B}^{\pi}$\\
\begin{figure}[!h]
		\centering
		\includegraphics[width=10cm]{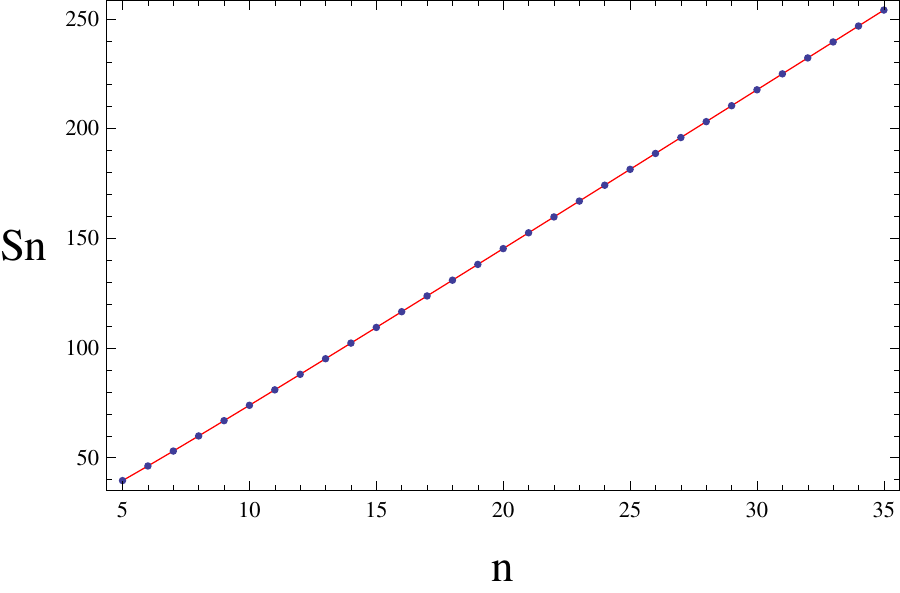}
		\caption{The entanglement entropy of gauge field with $\pi$ center.}
		\label{matterPicenter}
\end{figure}\\
The entanglement entropy with algebra $\mathcal{B}^{\pi}$ is shown in figure (\ref{matterPicenter}). We have the coefficients
\be
c_0=5.85105,\quad c_1=7.35487,\quad c_{\log}=-2.65376,\quad c_2=6.02547,\quad c_3=0.676344.
\ee
We can see that the logarithmic term is independent of the algebra choices.
\section{Mutual Information of 2D lattice gauge field with Coulomb gauge}
Now let us calculate the mutual information of gauge fields with Coulomb gauge between two squares of equal size $n^2$ separated by a distance $n$, for the four different algebra choices in figure (\ref{coulombtrivial}) and figure (\ref{coulombcenter}). The mutual information between region $A$ and $B$ is given by
\be
I(A,B)=S(A)+S(B)-S(A\cup B),
\ee
which is finite and well defined. \\

In figure (\ref{coulombtrivial}) and figure (\ref{coulombcenter}), we have the four algebras
\be
\mathcal{A}\supset \hat{\mathcal{A}}\supset \mathcal{A}^{\hat{A}},\mathcal{A}^{\pi}.
\ee
The mutual information is monotonously increasing with the algebra. Hence, we expect to have
\be\label{ordercoulomb}
I_{\mathcal{A}}(A,B)\geq I_{\hat{\mathcal{A}}}(A,B)\geq I_{\mathcal{A}^{\hat{A}}}(A,B),I_{\mathcal{A}^{\pi}}(A,B),
\ee
where $A$ and $B$ are the two square regions.\\

\begin{figure}[!h]
		\centering
		\includegraphics[width=10cm]{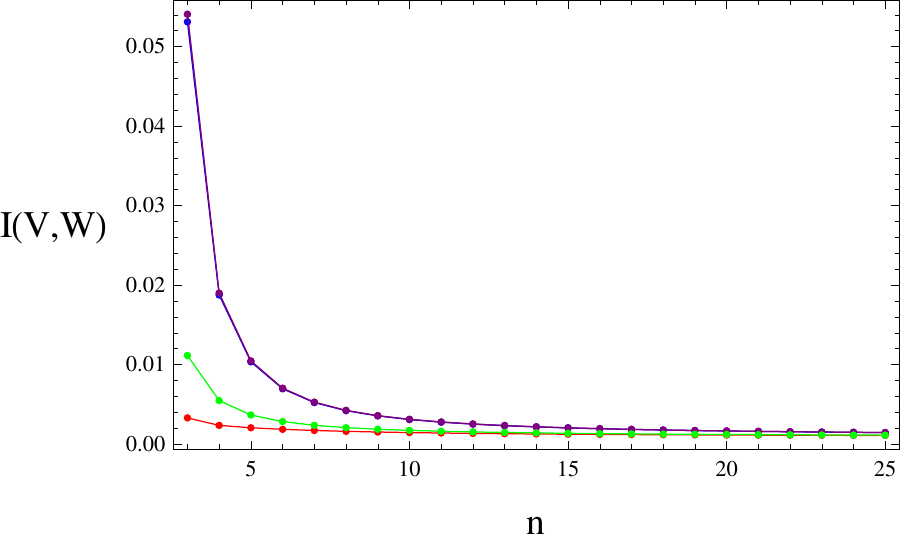}
		\caption{The mutual information of gauge fields with Coulomb gauge between two square regions of size $n$ and separation distance $n$ for four different algebra choices, from top to bottom $\mathcal{A}$ and $\hat{\mathcal{A}}$ (purple curve), $\mathcal{A}^{\hat{A}}$ (red curve) and $\mathcal{A}^\pi$ center (green curve). There are two curves in the purple curve. One is mutual information of algebra $\mathcal{A}$ and the other is of algebra $\hat{\mathcal{A}}$. At $n=3$, the mutual information of algebra $\mathcal{A}$ is a little larger than algebra $\hat{\mathcal{A}}$. For other values of $n$, they are hard to distinguish.}
		\label{mutualCoulomb}
\end{figure}

In figure (\ref{mutualCoulomb}), we show the numerical calculation of mutual information of gauge fields with Coulomb gauge between two square regions of the same area $n^2$ and separated by a distance $n$ for different algebra choices. The figure shows that the relation (\ref{ordercoulomb}) holds, as we expect. The mutual information for algebra $\mathcal{A}$ and $\hat{\mathcal{A}}$ is difficult to distinguish when $n\geq4$. When the regions become large, the effect of one pair of operators on the entropy and mutual information is negligible. This is because the algebra $\hat{\mathcal{A}}$ gets closer and closer to algebra $\mathcal{A}$, as the regions become larger and larger.

\section{Mutual Information of 2D lattice gauge field coupling with matter}
Now we consider the mutual information of gauge fields coupling with matter between two squares as in the previous section, again for the four different algebra choices in figure (\ref{matterphysical}). In figure (\ref{matterphysical}), we have the four algebras
\be
\mathcal{B}\supset \tilde{\mathcal{B}}\supset \mathcal{B}^{\tilde{A}},\mathcal{B}^{\pi}.
\ee
The mutual information is monotonously increasing with the algebra. Hence, we expect to have
\be\label{ordermatter}
I_{\mathcal{B}}(V,W)\geq I_{\tilde{\mathcal{B}}}(V,W)\geq I_{\mathcal{B}^{\tilde{A}}}(V,W),I_{\mathcal{B}^{\pi}}(V,W),
\ee
where $V$ and $W$ are the two square regions.\\

\begin{figure}[!h]
		\centering
		\includegraphics[width=10cm]{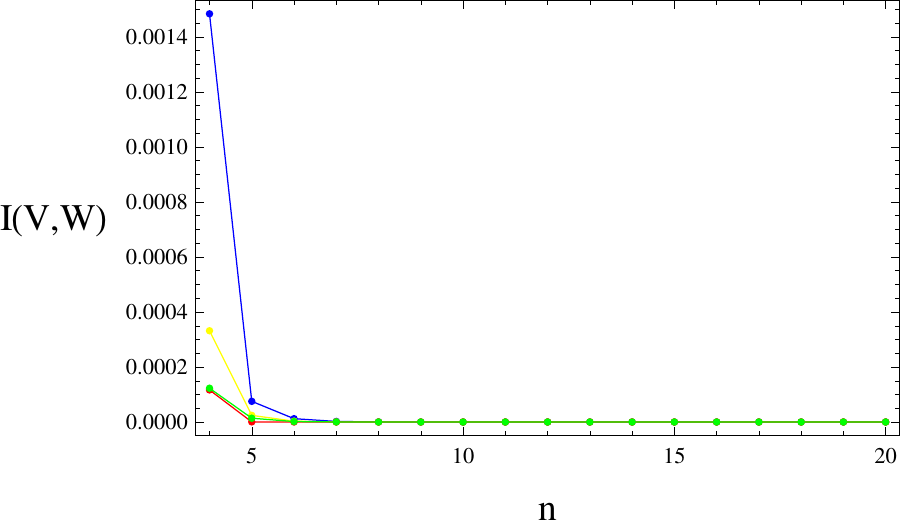}
		\caption{The mutual information of gauge fields coupling with matter between two squares of equal sizes separated by a distance equal to the square size for algebra $\mathcal{B}$ (blue curve), $\tilde{\mathcal{B}}$ (yellow curve), $\mathcal{B}^{\tilde{A}}$ (red curve) and $\mathcal{B}^\pi$ (green curve). We take $m=1$ in this figure. In the limit $n \rightarrow {\infty}$, the mutual information for all four algebra choices goes to $0$ very fast.}
		\label{mutualmatter}
\end{figure}

In figure (\ref{mutualmatter}), we take $m=1$ in the calculation. We show the numerical calculation of mutual information of gauge fields coupling with matter between two square regions of the same size $n$ and separated by a distance $n$ for different algebra choices. The figure shows that the relation (\ref{ordermatter}) holds, as we expect. The mutual information of all four different algebra choices decays to $0$ very quickly. This is one of the properties of gauge fields in the Higgs phase, the decay reflecting the mass gained.\\
\\

\begin{figure}[!h]
		\centering
		\includegraphics[width=10cm]{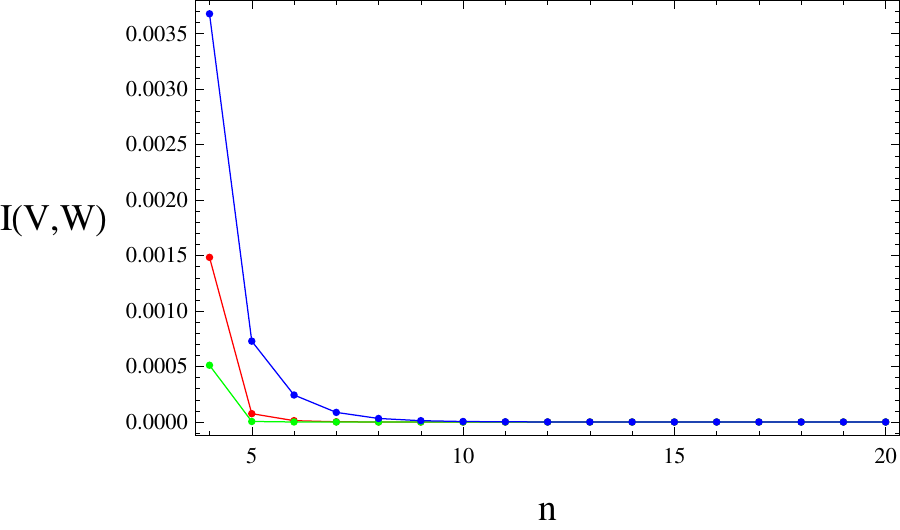}
		\caption{The mutual information of gauge fields coupling with matter between two squares of equal sizes separated by a distance equal to the square size for algebra $\mathcal{B}$ with $m=\frac{1}{2}$ (blue line),$m=1$ (red line) and $m=2$ (green line).}
		\label{mutualtrivialmatter}
\end{figure}

\begin{figure}[!h]
		\centering
		\includegraphics[width=10cm]{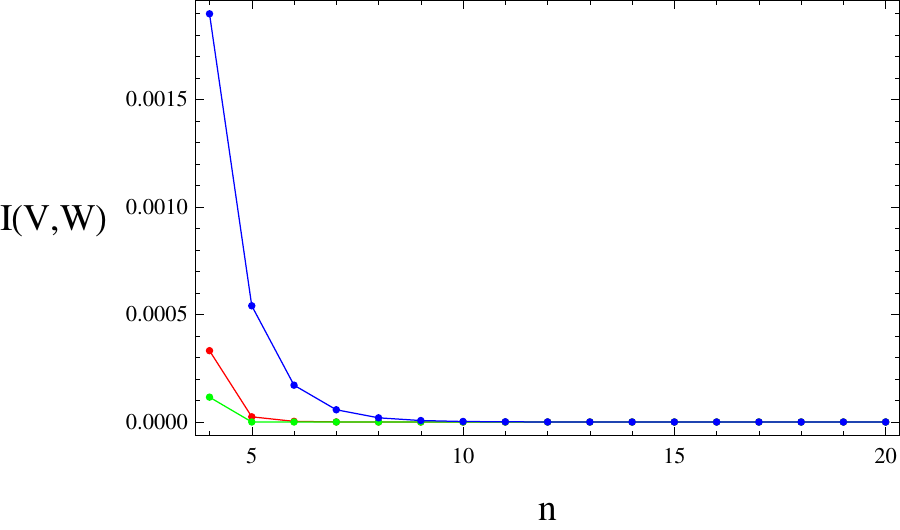}
		\caption{The mutual information of gauge fields coupling with matter between two squares of equal sizes separated by a distance equal to the square size for algebra $\tilde{\mathcal{B}}$ with $m=\frac{1}{2}$ (blue line),$m=1$ (red line) and $m=2$ (green line).}
		\label{mutualtrivialmatterdofremoved}
\end{figure}

\begin{figure}[!h]
		\centering
		\includegraphics[width=10cm]{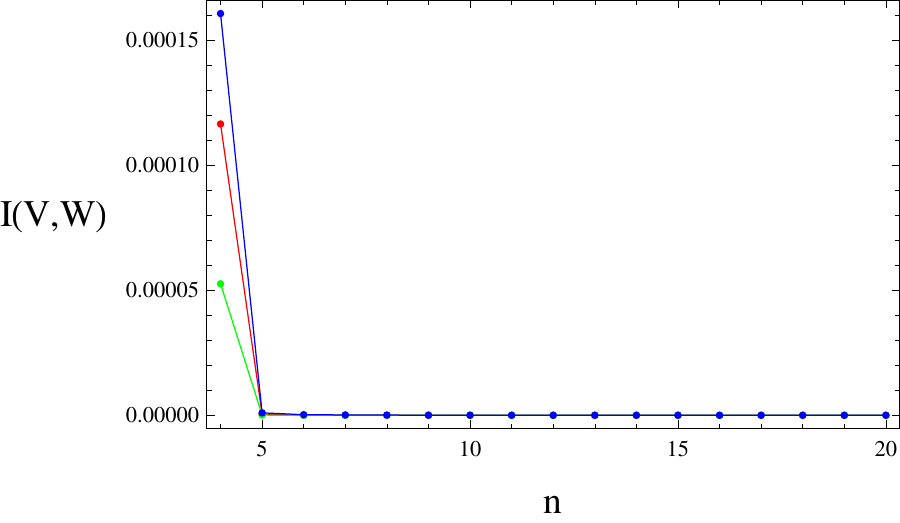}
		\caption{The mutual information of gauge fields coupling with matter between two squares of equal sizes separated by a distance equal to the square size for algebra $\mathcal{B}^{\tilde{A}}$ with $m=\frac{1}{2}$ (blue line),$m=1$ (red line) and $m=2$ (green line).}
		\label{mutualAmatter}
\end{figure}

\begin{figure}[!h]
		\centering
		\includegraphics[width=10cm]{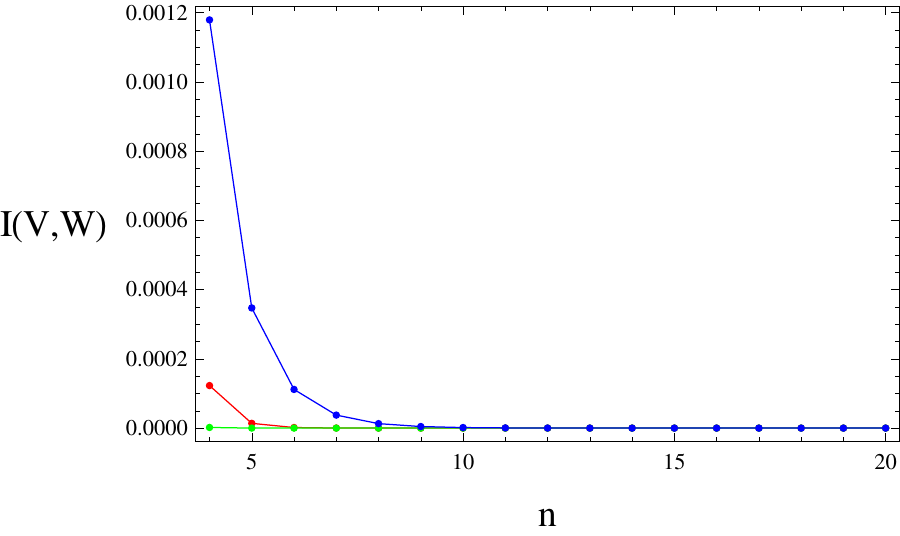}
		\caption{The mutual information of gauge fields coupling with matter between two squares of equal sizes separated by a distance equal to the square size for algebra $\mathcal{B}^\pi$ center with $m=\frac{1}{2}$ (blue line),$m=1$ (red line) and $m=2$ (green line).}
		\label{mutualpimatter}
\end{figure}

Now we compare the mutual information of gauge fields coupling with matter of different masses. We consider the same algebra choices and regions as the above case of $m=1$. As the mass of matter increases, the vacuum correlation functions of $\tilde{A}$ and $\pi$ decays faster with distance. We expect that the larger the mass, the smaller the mutual information. We compare the results of $m=\frac{1}{2},1,2$ for four different algebra choices in figure (\ref{mutualtrivialmatter}), figure (\ref{mutualtrivialmatterdofremoved}), figure (\ref{mutualAmatter}) and figure (\ref{mutualpimatter}) respectively.\\
\\
\\
\\
\\
Figure (\ref{mutualtrivialmatter}), figure (\ref{mutualtrivialmatterdofremoved}), figure (\ref{mutualAmatter}) and figure (\ref{mutualpimatter}) show the mutual information with different mass for four algebra choices $\mathcal{B}$, $\tilde{\mathcal{B}}$, $\mathcal{B}^{\tilde{A}}$ and $\mathcal{B}^{\pi}$ respectively. We can see that, for the four different algebra choices, the mutual information all decreases as the mass of matter increases.
\section{Conclusion}
In this paper, we have explored the effect of a different gauge choice on the tensor product structure of the Hilbert space constructed for the gauge potential in a U(1) gauge theory in 2+1 dimensions. In particular, departing from the usual choice of temporal gauge, we studied the Coulomb gauge and explicitly demonstrates that by imposing the constraints as second class constraints and obtaining Dirac brackets, the gauge potentials attain commutators that correspond to highly non-local operators-- commutators are non-vanishing all the way to infinite separation. To make sense of entanglement entropy associated to a local region, it is therefore necessary to construct a local operator algebra. We proposed a way making use of the duality with a scalar degree of freedom in the two dimensional spatial slice, in the cases of the massless theory and also in the Higgs phase. In the massless case, we recover the entanglement entropy of a scalar field (the log term in particular) and that of a U(1) theory when we pick a center. We also studied the case of the Higgs phase in the Coulomb gauge and construct a modified set of local operators accordingly. In this case, the choice of algebra has little detectable effect on the log term of the entanglement entropy. Our result highlights the fact that by choosing a different notion of fundamental operator basis, it naturally leads to a different set of algebra associated to a region.

We have shown, at least in 2+ 1 dimensions how to recover known results in terms of the gauge potential when the gauge potentials are non-local. This gives us some insights into the structure of the Hilbert space in a gauge theory. In terms of the gauge potentials, the Hilbert space is in some sense fluid -- local operators could turn into non-local ones based on different gauge choices.
However, suitable construction of local degrees of freedom appear to recover the expected result. This should have implications in gravitational theories as well.
We also note that our construction of a  set of local operators is based on a duality relation with scalars in 2 spatial dimensions. It would therefore be interesting to generalize our construction to higher dimensions.

We note that in this paper we have considered the Coulomb gauge and obtain Dirac brackets where the constraints are imposed explicitly at the level of canonical quantization. It is well known that in our usual path-integral quantization of the Maxwell theory, there are different gauge choices corresponding to extra term $\frac{1}{\zeta} \partial_\mu A^\mu$ in the action. One might wonder how different values of $\zeta$ is incorporated in the current discussion. We note that this term arises as we average over difference gauge choices $\partial A = \omega$ over a Gaussian distribution of $\omega$. This makes the discussion as some definite constraint in the context of canonical quantization at present unclear. We leave this interesting and important problem for future investigation.


\section*{Acknowledgements}
We thank Horacio Casini, Marina Huerta for their generous help with explaining the numerical method. We thank Qi Wang, Ce Shen and Xiangdong Zeng for help with coding.
LYH thank the support of Fudan University and the Thousands Young Talents Program. We also thank the referee specially for making very interesting comments about the generalized alpha gauge choices.


\begin{thebibliography}{99}

\bibitem{Buividovich:2008gq}
  P.~V.~Buividovich and M.~I.~Polikarpov,
  Phys.\ Lett.\ B {\bf 670}, 141 (2008)
  doi:10.1016/j.physletb.2008.10.032
  [arXiv:0806.3376 [hep-th]].

\bibitem{Casini:2013rba}
  H.~Casini, M.~Huerta and J.~A.~Rosabal,
  Phys.\ Rev.\ D {\bf 89}, no. 8, 085012 (2014)
  doi:10.1103/PhysRevD.89.085012
  [arXiv:1312.1183 [hep-th]].
\bibitem{Donnelly:2016auv}
  W.~Donnelly and L.~Freidel,
  JHEP {\bf 1609}, 102 (2016)
  doi:10.1007/JHEP09(2016)102
  [arXiv:1601.04744 [hep-th]].

\bibitem{Donnelly:2016rvo}
  W.~Donnelly and S.~B.~Giddings,
  Phys.\ Rev.\ D {\bf 94}, no. 10, 104038 (2016)
  doi:10.1103/PhysRevD.94.104038
  [arXiv:1607.01025 [hep-th]].

\bibitem{Donnelly:2014fua}
  W.~Donnelly and A.~C.~Wall,
  Phys.\ Rev.\ Lett.\  {\bf 114}, no. 11, 111603 (2015)
  doi:10.1103/PhysRevLett.114.111603
  [arXiv:1412.1895 [hep-th]].


\bibitem{Donnelly:2015hxa}
  W.~Donnelly and A.~C.~Wall,
  Phys.\ Rev.\ D {\bf 94}, no. 10, 104053 (2016)
  doi:10.1103/PhysRevD.94.104053
  [arXiv:1506.05792 [hep-th]].
\bibitem{Huang:2016bkp}
  X.~Huang and C.~T.~Ma,
  arXiv:1607.06750 [hep-th].

\bibitem{Ma:2015xes}
  C.~T.~Ma,
  JHEP {\bf 1601}, 070 (2016)
  doi:10.1007/JHEP01(2016)070
  [arXiv:1511.02671 [hep-th]].

\bibitem{Soni:2015yga}
  R.~M.~Soni and S.~P.~Trivedi,
  JHEP {\bf 1601}, 136 (2016)
  doi:10.1007/JHEP01(2016)136
  [arXiv:1510.07455 [hep-th]].

\bibitem{Ghosh:2015iwa}
  S.~Ghosh, R.~M.~Soni and S.~P.~Trivedi,
  JHEP {\bf 1509}, 069 (2015)
  doi:10.1007/JHEP09(2015)069
  [arXiv:1501.02593 [hep-th]].

\bibitem{Soni:2016ogt}
  R.~M.~Soni and S.~P.~Trivedi,
  JHEP {\bf 1702}, 101 (2017)
  doi:10.1007/JHEP02(2017)101
  [arXiv:1608.00353 [hep-th]].

\bibitem{Donnelly:2014gva}
  W.~Donnelly,
  Class.\ Quant.\ Grav.\  {\bf 31}, no. 21, 214003 (2014)
  doi:10.1088/0264-9381/31/21/214003
  [arXiv:1406.7304 [hep-th]].

\bibitem{Aoki:2015bsa}
  S.~Aoki, T.~Iritani, M.~Nozaki, T.~Numasawa, N.~Shiba and H.~Tasaki,
  JHEP {\bf 1506}, 187 (2015)
  doi:10.1007/JHEP06(2015)187
  [arXiv:1502.04267 [hep-th]].

\bibitem{Radicevic:2014kqa}
  D.~Radicevic,
  arXiv:1404.1391 [hep-th].

\bibitem{Gromov:2014kia}
  A.~Gromov and R.~A.~Santos,
  Phys.\ Lett.\ B {\bf 737}, 60 (2014)
  doi:10.1016/j.physletb.2014.08.023
  [arXiv:1403.5035 [hep-th]].

\bibitem{Hung:2015fla}
  L.~Y.~Hung and Y.~Wan,
  JHEP {\bf 1504}, 122 (2015)
  doi:10.1007/JHEP04(2015)122
  [arXiv:1501.04389 [hep-th]].

\bibitem{Radicevic:2015sza}
  Ð.~Radičević,
  JHEP {\bf 1604}, 163 (2016)
  doi:10.1007/JHEP04(2016)163
  [arXiv:1509.08478 [hep-th]].



\bibitem{Mathur:2015wba}
  M.~Mathur and T.~P.~Sreeraj,
  Phys.\ Rev.\ D {\bf 92}, no. 12, 125018 (2015)
  doi:10.1103/PhysRevD.92.125018
  [arXiv:1509.04033 [hep-lat]].

\bibitem{VanAcoleyen:2015ccp}
  K.~Van Acoleyen, N.~Bultinck, J.~Haegeman, M.~Marien, V.~B.~Scholz and F.~Verstraete,
  Phys.\ Rev.\ Lett.\  {\bf 117}, no. 13, 131602 (2016)
  doi:10.1103/PhysRevLett.117.131602
  [arXiv:1511.04369 [quant-ph]].

\bibitem{Casini:2014aia}
  H.~Casini and M.~Huerta,
  Phys.\ Rev.\ D {\bf 90}, no. 10, 105013 (2014)
  doi:10.1103/PhysRevD.90.105013
  [arXiv:1406.2991 [hep-th]].

\bibitem{Aoki:2017ntc}
  S.~Aoki, N.~Iizuka, K.~Tamaoka and T.~Yokoya,
  Phys.\ Rev.\ D {\bf 96}, no. 4, 045020 (2017)
  doi:10.1103/PhysRevD.96.045020
  [arXiv:1705.01549 [hep-th]].

\bibitem{EE}
  M. Ohya and D. Petz, ��Quantum entropy and its use��, Text and Monographs in Physics, Springer Study Edition, Corrected 2nd Printing, 2004.
\bibitem{Weinberg}
  S. Weinberg ��The Quantum Theory of Fields��,Volume 1,Cambridge University Press, 2005.


\end{thebibliography}
\end{document}